\documentclass[superscriptaddress,nofootinbib,tightenlines,preprint]{revtex4-1}
\pdfoutput=1
\usepackage[utf8]{inputenc}

\usepackage[pdftex,breaklinks,colorlinks,linktocpage,
citecolor=blue,
urlcolor=blue]{hyperref}

\usepackage{bm}
\usepackage{amssymb}
\usepackage{amsfonts}
\usepackage{mathrsfs}
\usepackage{amsmath}
\usepackage{amsthm}
\usepackage{xcolor}
\usepackage{bbold}
\usepackage{braket}
\usepackage{physics}
\usepackage{graphicx}
\usepackage{float}
\usepackage{caption}
\usepackage{subcaption}
\usepackage{url}
\usepackage{cancel}
\usepackage{soul}
\usepackage[normalem]{ulem}
\usepackage{enumitem}

\usepackage{multirow}
\usepackage{rotating}
\usepackage{tikz}
\usetikzlibrary{decorations.pathmorphing} % Needed for zigzag

\newcommand{\be}{\begin{equation}}
\newcommand{\ee}{\end{equation}}
\newcommand{\beq}{\begin{equation}}
\newcommand{\eeq}{\end{equation}}
\newcommand{\bea}{\begin{eqnarray}}
\newcommand{\eea}{\end{eqnarray}}
\def\bml{\begin{subequations}}
\def\blea{\bml\begin{eqnarray}}
\def\eml{\end{subequations}}
\def\elea{\end{eqnarray}\eml}
\newcommand{\e}{\mathrm{e}}

\definecolor{darkgreen}{RGB}{34,140,34}

\begin{document}

\title{Mapping 1+1-dimensional black hole thermodynamics to finite volume effects}

\author{Jean Alexandre}
\email{jean.alexandre@kcl.ac.uk}
\affiliation{Department of Physics, King’s College London, Strand, London, WC2R 2LS, United Kingdom}

\author{Drew Backhouse}
\email{drew.backhouse@kcl.ac.uk}
\affiliation{Department of Physics, King’s College London, Strand, London, WC2R 2LS, United Kingdom}

\author{Eleni-Alexandra Kontou}
\email{eleni.kontou@kcl.ac.uk}
\affiliation{Department of Mathematics, King’s College London,
Strand, London WC2R 2LS, United Kingdom}

\author{Diego Pardo Santos}
\email{diego.pardo@kcl.ac.uk}
\affiliation{Department of Mathematics, King’s College London,
Strand, London WC2R 2LS, United Kingdom}

\author{Silvia Pla}
\email{silvia.pla\_garcia@kcl.ac.uk}
\affiliation{Department of Physics, King’s College London, Strand, London, WC2R 2LS, United Kingdom}

\begin{abstract}

Both black hole thermodynamics and finite volume effects in quantum field theory violate the null energy condition.
Motivated by this, we compare thermodynamic features between two $1+1$-dimensional systems: 
{\it(i)} a scalar field confined to a periodic spatial interval of length $a$ and tunneling between two degenerate vacua;
{\it(ii)} a dilatonic black hole at temperature $T$ in the presence of matter fields. If we identify $a\propto T^{-1}$, we find similar thermodynamic behaviour,
which suggests some deeper connection arising from the presence of non-trivial boundary conditions in both systems. 
 We then extend our results to $2+1$ and $3+1$-dimensions and,
although a more complete study is necessary, the connection found in $1+1$-dimensions seems to be valid in higher dimensions too.

\end{abstract}

\begin{flushright}
\raggedleft
    \small KCL-PH-TH/2024-29
\end{flushright}

\maketitle

\tableofcontents

\newpage

\section{Introduction}

One of the seminal results in semi-classical gravity is Hawking radiation, and subsequent black hole evaporation \cite{Hawking:1974rv}. Part of its importance is that it shows a clear quantum effect in a regime where the curvature is small (the black hole horizon) and thus it can be treated classically. An important consequence of Hawking radiation is the violation of the null energy condition (NEC) around the black hole horizon.

The NEC is part of the classical energy conditions which are what we call \textit{pointwise}: they restrict some contraction of the stress tensor at every spacetime point (see \cite{Curiel:2014zba} and \cite{Kontou:2020bta} for reviews). The NEC is the weakest of them and it is written as
\be
T_{\mu \nu} \ell^\mu \ell^\nu \geq 0 \,,
\ee
where $T_{\mu \nu}$ is the stress-energy tensor and $\ell^\mu$ a null vector field. The NEC is obeyed by most classical fields\footnote{It is however violated by scalars with non-minimal coupling to gravity \cite{Barcelo:2000zf, Fliss:2023rzi}.} and it is often considered an important property of physical matter. Its geometric form, obtained by the use of the Einstein equation, is called the null convergence condition and it implies that a non-rotating null geodesic congruence locally converges. It was famously used in Penrose's singularity theorem \cite{Penrose:1964wq}, Hawking's black hole area theorem \cite{Hawking:1971vc} and other classical relativity results. If the stress-energy tensor has the form of a perfect fluid, $T_{\mu \nu}=(\rho +p) v^\mu v^\nu+pg^{\mu \nu}$, where $\rho$ is the energy density, $p$ the pressure and $v^\mu$ is the fluid’s unit four-velocity vector field, the NEC becomes
\be\label{NECpf}
\rho+p \geq 0 \,.
\ee
The NEC, as is the case for all pointwise energy conditions, is violated in the context of semi-classical gravity; with the most prominent case being the Hawking radiation. More generally, quantum field theories (QFTs) obeying some reasonable axioms always have states that admit negative energy as shown by Epstein, Glaser and Jaffe in the 1960's \cite{Epstein:1965zza}. 

Interestingly, the NEC is also violated in a different setting, involving finite volume effects in scalar QFT. A finite volume in QFT implies two fundamental features: quantisation of momentum and tunnelling between multiple vacua.
The first feature is at the origin of the Casimir effect (see \cite{Bordag} for a review), which is known to induce NEC violation. As shown more recently and reviewed in the next section, the second feature also leads to NEC violation, in relation to convexity of the effective potential \cite{Alexandre:2022qxc,Alexandre:2023iig,Alexandre:2023pkk,Alexandre:2023bih,Ai:2024taz}. For both the Casimir effect and tunnelling, NEC violation arises from a ground state energy which is not extensive, i.e. not simply proportional to the size of the system.

We investigate the possibility of a correspondence between these two sources of NEC violation. In particular:
{\it(i)} a scalar field confined to a periodic spatial interval of length $a$ and tunnelling between two degenerate vacua in the limit of zero temperature;
{\it(ii)} a dilatonic black hole at temperature $T$ in the presence of matter fields in an infinite spatial volume. For simplicity, we focus mainly on these two $1+1$-dimensional systems although we briefly discuss $2+1$ and $3+1$-dimensional systems in section \ref{sec:higherd}.

The motivation is to find common features between two non-trivial thermodynamical systems due to quantum effects. Our main results are summarised in table \ref{Tab:Summary}, in the regime $ma\propto M/T\lesssim1$, where $m,a$ are respectively the mass and length scales in the tunnelling description, and $M,T$ are respectively the mass and temperature of the dilatonic black hole (which are independent parameters).
The two systems are thermodynamically similar under the matching condition of $a\propto 1/T$, suggesting a mapping between finite size and finite temperature.
It is important to note that this analogy cannot be attributed to dimensional considerations though, since several length/mass scales are present in both models.

We stress here that this work does not establish rigorously a duality between the two systems, 
as one could hope such as the AdS/CFT correspondence for example. The aim of this approach is to put forward a complementary study, 
which could provide a different angle on black hole thermodynamics, based on an analogy with a simpler system in flat spacetime. 
Our strategy is to first derive new properties, both for the confined scalar field and the dilatonic black hole. The resulting mapping $T\leftrightarrow a^{-1}$ we then find is not trivial, and suggests that further studies could be made,
requiring a more systematic formalism. Our results are therefore preliminary, but promising for a new and original mapping.

\begin{table}
	\centering
	\begin{tabular}{ccc}
		\hline\\[-10pt]
		\quad System \quad &
		\quad NEC violation \quad &
		\quad Entropy rate \quad \\[5pt]
		\hline
		\hline \\[-5pt]
		Tunneling in 1+1 flat spacetime &
		$-\dfrac{\pi}{3a^2}$ &
		$\dfrac{1}{2} -\dfrac{2m}{3}a$ \\[20pt]
		1+1 Dilatonic Black Holes &
		$-\dfrac{N \pi^2}{12} T^2$ &
		$-\dfrac{N}{12} -\dfrac{M}{\pi T}$ \\[15pt]
		\hline
	\end{tabular}
	\caption{\small{Summary of results comparing the two thermodynamical systems, for NEC violation and the rate of change in entropy:
			{\it(i)} Tunneling in 1+1 flat spacetime (cf. eqs. \eqref{eq:necTmal1} and \eqref{Lengthrate});
			{\it(ii)} 1+1 Dilatonic Black Holes (cf. eqs. \eqref{NECQuantumBH} and \eqref{Temperaturerate}). $N$ is the number of massless scalar fields.}}
	\label{Tab:Summary}
\end{table}

In section~\ref{sec:Euclidean} we calculate the free energy for the ground state resulting from the scalar field tunnelling between degenerate minima.
We explain why in one space dimension, the Casimir and tunnelling effects are of the same order of magnitude. In the limit of vanishing temperature, NEC violation can be decomposed as the sum of two contributions: one from the Casimir effect and one from tunnelling. We show that the latter is actually more important than the former if $ma\gtrsim1$, which is a new feature with relevance potentially going beyond the present study.

Section~\ref{sec:dilatonicBHs} presents the derivation of the thermodynamical properties of the dilatonic black hole in the presence of non-self-interacting matter fields. A detailed explanation is given for the role of the environment regarding the entropy of the system once backreaction of the matter fields on the background metric is taken into account. 

In section~\ref{sec:comparison} we compare the two studies and we find that the relation $a\propto T^{-1}$ provides a mapping between the two systems.

 Finally, section \ref{sec:higherd} extends our results to $2+1$ and $3+1$-dimensions, where some results are known and the comparison still holds for certain cases. Higher dimensions allow for different geometries/topologies for the confining space in flat spacetime, but also more black hole solutions, with different asymptotic behaviours.
We find in 2+1 dimensions that both the NEC violation and the entropy follow the same analogy as the one we have in 1+1 dimensions. 
In 3+1 dimensions the analogy for the NEC violation clearly holds in specific regimes, but more work needs to be done as far as the entropy is concerned.

Tunnelling at finite temperature is described with a Euclidean metric whereas the metric sign convention for the black hole description is $(-,+)$. 
Natural units of $\hbar=c=1$ are used throughout.

\section{Finite size effects in 1+1 dimensional flat spacetime}
\label{sec:Euclidean}

We consider a massive self-interacting real scalar field theory defined on a one-dimensional periodic interval $x\in[0,a]$ at a temperature $T\equiv1/\beta$ with a corresponding Euclidean action
\be
\label{Sbare}
I=\frac{1}{2}\int_{0}^{\beta}\text{d}\tau\int_{0}^{a} \text{d}x
\left((\dot\phi)^2+(\phi')^2+\frac{m^2}{4}(\phi^2-1)^2\right)~,
\ee
where $m$ is the mass scale of the theory. Starting from this we study the thermodynamics of the true vacuum of the effective theory.

\subsection{Convexity from tunnelling}
\label{sec:convexity}

\begin{figure}
	\centering
	\includegraphics[width=0.6\linewidth]{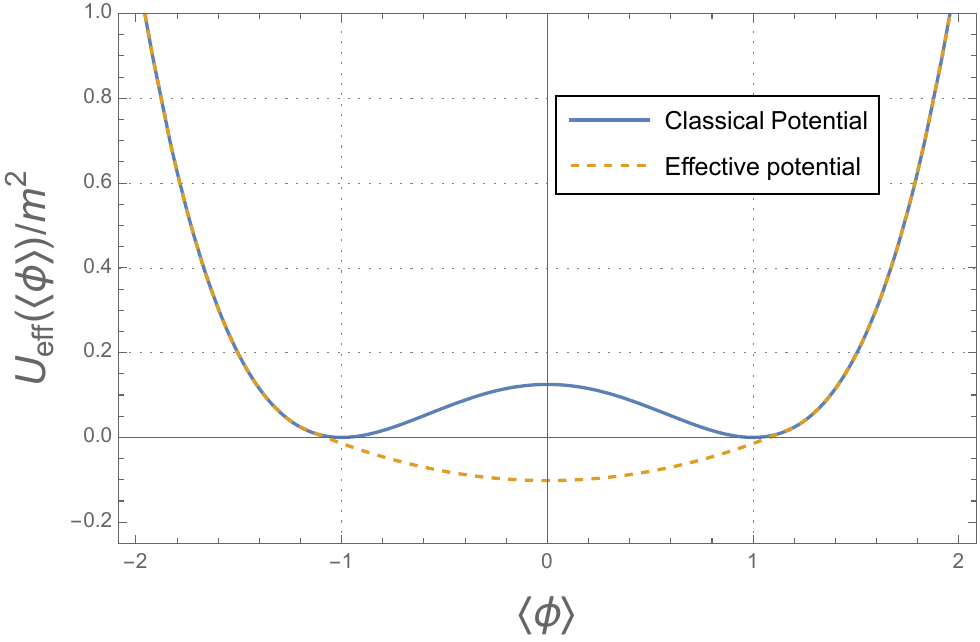}
	\caption{\small{The classical potential (solid blue) and effective potential (dashed orange) in a finite spatial volume. The effective potential is necessarily convex due to tunneling between the two degenerate vacua, restoring symmetry.}}
	\label{fig:Convexity}
\end{figure}

It is known that the one-Particle-Irreducible (1PI) effective potential is necessarily convex if one takes several vacua into account 
\cite{Symanzik:1969ek, Coleman:1974jh, Iliopoulos:1974ur, Haymaker:1983xk, Fujimoto:1982tc, Bender:1983nc, Hindmarsh:1985nc, Alexandre:2012ht, Plascencia:2015pga, Millington:2019nkw}. Focusing on two degenerate vacua at $\pm v$,
the dynamics of this feature relies on tunnelling between these vacua \cite{Alexandre:2022qxc}, which restores symmetry and 
induces a true ground state corresponding to a vanishing field expectation value $\left<\phi\right>=0$ (see figure~\ref{fig:Convexity}).
Equivalently, by symmetry of the bare potential, the true vacuum corresponds to a vanishing source $j=0$, and
\be
\left<\phi\right>\equiv-\frac{1}{Z[j]}\left.\frac{\delta Z[j]}{\delta j}\right|_{j=0}=0~,
\ee
where $Z[j]$ is the partition function.
The picture described here can be interpreted as backreaction: the double-well bare potential allows quantum fluctuations to tunnel
between the minima, which in turns modifies the vacuum structure by imposing convexity. The resulting symmetric vacuum corresponds then to a
non-perturbative process, which can be described by the semi-classical approximation for $Z[j]$, as explained below.

Symmetry restoration is possible in a finite volume only, though, since an infinite volume implies Spontaneous Symmetry Breaking instead. But a finite volume requires a discrete set of momenta for quantum fluctuations and, 
as we show in this article, in the limit of vanishing temperature, finite-size effects can be decomposed as the sum of two contributions: 
{\it(i)} discretisation of momentum for quantum fluctuations, that we will refer to as the Casimir effect;
{\it(ii)} symmetry restoration due to tunnelling, that we will refer to as the tunnelling effect.

Allowing tunnelling between two degenerate vacua, the 1PI effective potential induced by a dilute gas of instantons was calculated in
\cite{Alexandre:2023iig}, based on an expansion in $\left<\phi\right>$ to the quadratic order. 
This result explicitly shows a convex effective potential, with a positive mass term and a true vacuum at $\left<\phi\right>=0$. 
Focusing on this true vacuum, the complete one-loop quantisation of the dilute gas with discrete momentum is calculated in \cite{Ai:2024taz}
for a 3-torus, providing the full picture of the interplay between Casimir and tunnelling effects. 

These calculations were done in 3+1 dimensions though, and we consider here the 1+1 dimensional case, where both effects are comparable. 
Indeed, for a finite length $a$ and a mass scale $m$, the instanton action is of the order $ma$, leading for $ma\gg1$ to a suppression of the tunnelling 
effect of the order $\exp(-ma)$, similarly to what happens in the Casimir effect.\footnote{Other studies of finite volume QFT effects in 1+1 dimensional spacetime can be found in \cite{Hu:2021fjq}.}

As we show in this section, one feature of convexity obtained from quantum fluctuations is NEC violation in the true ground state. 
If we allowed the system to evolve freely, 
this violation would imply an increase in the length $a$, similarly to spacetime expansion due to tunnelling-induced NEC violation, 
as described in \cite{Alexandre:2023pkk, Alexandre:2023bih} 
(see for example \cite{Rubakov:2014jja, Easson:2024fzn} for reviews of NEC violation in the context of Cosmology).
In the present work, we do not take into account spacetime dynamics though, 
and we stick to static QFT. This implies that some external system fixes the length $a$, which requires some energy. 
As a consequence, although the confined scalar field violates the NEC, the Averaged NEC is not violated, which can be seen by integrating the NEC along a null geodesic
going through the confining walls \cite{Sopova:2002cs, Graham:2002yr, Graham:2005cq}.

The main part of this section focuses on the zero-temperature case, since thermal effects tend to restore the NEC. 
Nevertheless we start the calculations at finite temperature $\beta^{-1}$, and consider then the limit $\beta\to\infty$.

\subsection{Tunneling and the Casimir effect}

\subsubsection{Semi-classical approximation and true vacuum}

Defining the dimensionless variables
\be
t \equiv m\tau~~~~\mbox{and}~~~r \equiv m x~,
\ee
the action of eq.~\eqref{Sbare} becomes
\be
I=\frac{1}{2}\int_{0}^{m\beta}\text{d}t\int_{0}^{ma} \text{d}r\left((\dot\phi)^2+(\phi')^2+\frac{1}{4}(\phi^2-1)^2\right)~,
\ee 
where dots and primes now represent derivatives in $t$ and $r$ respectively. One can see that this action 
depends on the two dimensionless parameters $ma$ and $m\beta$, and is invariant under the simultaneous rescaling
\be
a\to\lambda a~~,~~\beta\to\lambda\beta~~,~~m\to m/\lambda~,
\ee
and the quantum theory should also respect this symmetry, as we confirm in what follows. The equation of motion (EoM) with solutions $\phi_i$ is
\be \label{EoM}
\ddot\phi_i+\frac{1}{2}\left(\phi_i-\phi_i^3\right) =0~,
\ee
where in this work we consider only Euclidean-time-dependent and homogeneous configurations $\phi_i$,
since vacuum bubbles forming from degenerate vacua would have an infinite radius \cite{Coleman:1977py,Callan:1977pt}. There are several solutions to the classical EoM \eqref{EoM}, each to be studied in following subsections.

For a vanishing source $j=0$ and in the semi-classical approximation, the partition function $Z[0]\equiv Z$
can be approximated as a sum of path integrals over regions in field space around the action minimising saddle points $\varphi_i$ via the field decomposition $\varphi=\varphi_i+\psi$, integrating over fluctuations $\psi$.
This assumes that the fluctuations do not overlap,
and the one-loop approximation for the fluctuation factors leads to
\bea \label{Zsc}
Z&=&\int\mathcal{D}[\phi]
~\exp\Big(-I[\phi]\Big)\\
&\simeq& \sum_i\int\mathcal{D}[\psi] ~\exp\Big(-I[\varphi_i+\psi]\Big)\nonumber\\
&=&\sum_i \Big(\mbox{det}(\delta^2I[\phi_i])\Big)^{-1/2}
\exp\Big(-I[\phi_i]\Big)\nonumber\\
&\equiv&\sum_i \exp\Big(- W[\phi_i]\Big)~,\nonumber
\eea 
where the individual connected graph generating functionals are
\be \label{eqn:W}
W[\phi_i]\equiv I[\phi_i]
+\frac{1}{2}\Tr
\left\{\ln\left(\delta^2I[\phi_i]\right)\right\}~.
\ee

\subsubsection{Static saddle points}

There are two static saddle points in the present model, $\phi_s=\pm1$.
The corresponding individual connected graph generating functionals \eqref{eqn:W} can be evaluated using known methods developed for the study of the thermal Casimir effect on a 1D periodic interval \cite{Bordag}. The steps are outlined in appendix \ref{Casimir} and lead to
\be \label{Wstat}
W_\text{stat}(a,\beta)
\equiv W[\phi_s]
=a\beta \Lambda^2
-\frac{m^2a\beta}{\pi}\int_1^\infty \text{d}u\frac{\sqrt{u^2-1}}{e^{mau}-1}
+\sum_{n\in\mathbb{Z}}\ln\left(1-\e^{-\beta\omega_n}\right)~,
\ee
where $\Lambda^2$ is an ultraviolet cutoff,
corresponding to the vacuum energy of unbounded space,
and the quantised frequencies/wave vectors are
\be
\omega_n=\sqrt{m^2+k_n^2}
\quad,\quad
k_n=\frac{2\pi n}{a}~.
\ee
We note here that quantum corrections indeed depend on $a,\beta,m$ through the products $ma$ and $m\beta$ only.

\subsubsection{Time-dependent saddle points}
\label{Spacetime-dependent Saddle points}

The fundamental time-dependent saddle point is the (anti-)instanton relating the two vacua of the bare potential 
\be 
\phi_\text{inst}(t)=(\pm)\tanh \left(\frac{t -\text{$t_1 $}}{2}\right)~,\label{eqn:Instanton}
\ee
where $t_1$ is the time of the jump.
At a finite temperature, field configurations are periodic in Euclidean time and hence instantons and anti-instantons can only exist in pairs.
Such field configurations are well approximated by a product of individual (anti-)instanton configurations
\be
\phi_\text{n-pair}(\tau)\simeq\prod_{j=1}^{2n}(-1)^j\tanh\left(\frac{t-t_j}{2}\right)~,
\ee
provided that the jumps at $t_i$ and $t_j$ are sufficiently distant $(|t_i-t_j|\gg1)$.
The factor $-1$ ensures that an instanton is always followed by an anti-instanton and the product is taken to $2n$ to enforce periodicity in Euclidean time.
In the limit of small temperature, $m\beta\gg1$, a large amount of instanton/anti-instanton pairs is allowed and we assume in what follows the 
instanton dilute gas approximation \cite{Kleinert:2004ev}, where the width of each jump is negligible compared to $m\beta$. Also,
(anti-)instantons are far enough from each other for them to keep their shape, which for $n$ pairs leads to the total action
\be
I_\text{n-pairs}\simeq 2n I_\text{inst}~,
\ee
where the action for one (anti-)instanton is
\be \label{Instaction}
I_\text{inst}\equiv I[\phi_\text{inst}]=\frac{2ma}{3}~.
\ee
The fluctuation factor for $n$ instanton/anti-instanton pairs can then be approximated by the product of 
fluctuation factors for each static saddle point evaluated over half the total Euclidean time interval $\beta/2$, times
the fluctuation factors for each instanton jump. The corresponding connected graph generating functional is then
\be
W_\text{n-pairs}
\equiv W[\phi_\text{n-pair}]
\simeq 2W_\text{stat}(a,\beta/2)+2nW_\text{jump} \,,
\ee 
where we know from tunnelling in Quantum Mechanics \cite{Kleinert:2004ev} that
\be 
W_\text{jump}\equiv I_\text{inst}-\frac{1}{2}\ln(\frac{6 I_\text{inst}}{  \pi })~.
\ee
We note that the expression for $W_\text{jump}$ takes into account time-dependent quantum fluctuations over the instantons, and it neglects the space-dependence of these fluctuations.
However, it was shown in \cite{Ai:2024taz} that the main contribution of the instanton jump comes from the zero-modes, validating the approximation made here.

\subsubsection{Partition function}

Assuming the semi-classical approximation and a dilute gas of instantons/anti-instantons, the partition function can be expressed as a sum over
all the possible $n$-pair configurations
\be \label{Zsemi-classicalLowT}
Z\simeq
2\exp\big(- W_\text{stat}(a,\beta)\big)
+2\sum_{n=1}^\infty \left(\prod_{i=1}^{2n}\int_{t_{i-1}}^{m\beta}\mathrm{d}t_i\right) 
\exp(-2 W_\text{stat}(a,\beta/2)-2n W_\text{jump})~.
\ee
The first term in the right-hand side corresponds to the two static saddle points.
In the second term the product of integrals accounts for the invariance of the total action $I_\text{n-pairs}$
under the translations of each successive instanton jump over the remaining dimensionless Euclidean time interval $t\in[t_i,m\beta]$, defining $t_0\equiv0$ since the first instanton can exist over the whole interval. 
This invariance under translation of the jumps corresponds to the zero modes of the fluctuation factors for each $n-$pair configuration.
The factor 2 takes into account the instanton configurations starting and ending at either $+1$ or $-1$. Using the known result \cite{Coleman:1977py}
\be 
\prod_{i=1}^{2n}\int_{t_{i-1}}^{m\beta}\mathrm{d}t_i=\frac{(m\beta)^{2n}}{(2n)!}~,
\ee
the partition function \eqref{Zsemi-classicalLowT} can be expressed as
\be \label{Zsemi-classicalLowT2}
Z\simeq
2\exp\big(- W_\text{stat}(a,\beta)\big)
+2\exp\big(-2 W_\text{stat}(a,\beta/2)\big)\sum_{n=1}^\infty\frac{\bar{N}^{2n}}{(2n)!}~,
\ee
where
\be\label{Nbar}
\bar N\equiv m\beta \sqrt{\frac{6 I_\text{inst}}{  \pi}} e^{- I_\text{inst}}
=2m\beta\sqrt\frac{ma}{  \pi}~\exp\left(-\frac{2ma}{3}\right)~,
\ee 
with $\bar N/2$ corresponding to the average number of instanton/anti-instanton pairs over the whole Euclidean time $\beta$ \cite{Alexandre:2022qxc}. In the small temperature limit $(m\beta\gg1)$ the last term in $W_\text{stat}$ \eqref{Wstat} can be neglected, such that it can be taken as linear in $\beta$
\be\label{Wapprox}
W_\text{stat}(a,\beta)\simeq 2W_\text{stat}(a,\beta/2)~,
\ee
and the partition function \eqref{Zsemi-classicalLowT2} becomes
\footnote{The approximation \eqref{Wapprox} is applied to the static saddle point contribution, and not to the instanton contribution, since it is sub-dominant at low temperatures.}
\be \label{Zsemi-classicalLowT3}
Z\simeq
2\exp\big(-2 W_\text{stat}(a,\beta/2)\big)\sum_{n=0}^\infty\frac{\bar{N}^{2n}}{(2n)!}
=2\exp\big(-2 W_\text{stat}(a,\beta/2)\big)\cosh(\bar{N})~.
\ee
Finally, in the limit $m\beta\gg1$ for finite $ma$ (such that $\bar{N}\gg1$) the total free energy  is
\be \label{FDisTrue}
F_\text{true}\equiv-T\ln(Z)\simeq 2TW_\text{stat}(a,\beta/2)-T\bar{N}~,
\ee
and corresponds to the sum of the usual free-field Casimir contribution $2TW_\text{stat}(a,\beta/2)$ 
and the tunneling contribution $-T\bar{N}$.

\subsection{Null Energy Condition} \label{Sec:NEC}

We show here that the true ground state of the system we consider violates the NEC, as a consequence of the true vacuum energy not being extensive: 
the free energy (\ref{FDisTrue}) is not simply proportional to $a$.

We assume here that the dilute instanton gas described by the partition function \eqref{Zsemi-classicalLowT3} may be treated as a perfect fluid, such that the resulting null energy condition reduces to the simpler form \eqref{NECpf}. The thermodynamic energy density and pressure are then defined as
\bea
\rho &\equiv& \frac{1}{a}\left(F_\text{true}+\beta\frac{\partial F_\text{true}}{\partial\beta}\right),\\
p &\equiv& -\frac{\partial F_\text{true}}{\partial a} \,.
\eea
Here we show that the NEC is violated by the finite volume effects we consider.

\begin{figure}
	\centering
	\includegraphics[width=0.7\linewidth]{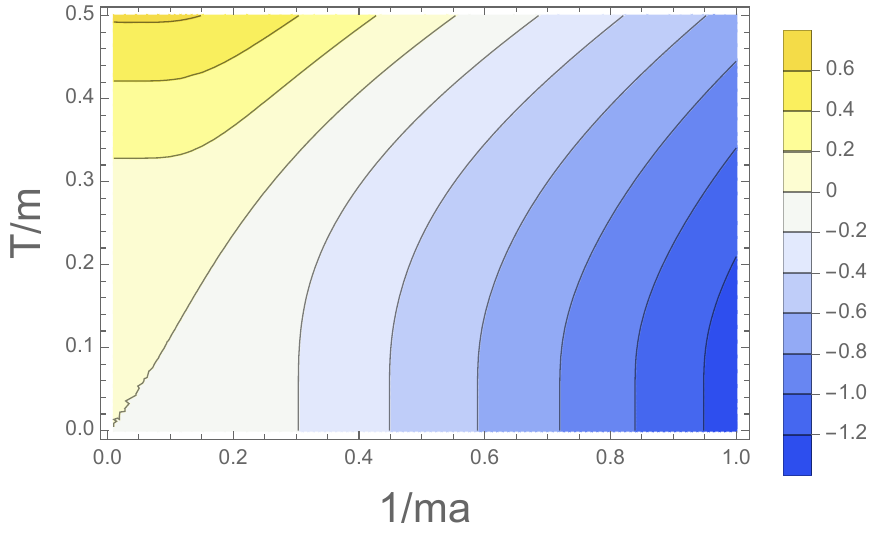}
	\caption{{\small Numerical plot of $(\rho+p)/m^2$ \eqref{rho+p} for inverse dimensionless length $1/ma$ with finite temperature corrections. At zero temperature, NEC violation increases as the spatial interval reduces, corresponding to an increased tunneling rate, and approaches zero in the limit of infinite spatial length where tunneling is completely suppressed. For a given length scale, finite temperature effects increase $\rho+p$ and can lead to NEC satisfaction at large length scales.}}
	\label{fig:NEC}
\end{figure}

The sum $\rho+p$ may be evaluated from the free energy \eqref{FDisTrue} and satisfies
\bea\label{rho+p}
\frac{\rho+p}{m^2} &=&
-\frac{ma}{\pi}\int_{1}^\infty
du\frac{u\e^{mau}\sqrt{u^2-1}}{\left(\e^{mau}-1\right)^2}
-\frac{4ma+3  }{3  \sqrt{\pi ma  }}\exp\left(-\frac{2ma}{3  }\right)\\
&&+\sum_{n\in\mathbb{Z}}\frac{m^2a^2+8n^2\pi^2}{m^2a^3\omega_n}\left(\e^{\beta\omega_n/2}-1\right)^{-1}~,\nonumber
\eea
where we can identify the following terms:
\begin{enumerate}[label=(\roman*)]
	\item  the first term (integral over $t$) corresponds to the known Casimir effect, 
	obtained for a free field, with a negative contribution; 
	\item  the second term corresponds to tunnelling arising from degenerate vacua, with a negative contribution;\
	\item  the third term (sum over Matsubara modes) corresponds to finite temperature effects providing a positive contribution, and becomes the usual contribution from black body radiation $(\propto T^2)$ in the massless and infinite length limit.
\end{enumerate}
The expression \eqref{rho+p} is plotted in figure~\ref{fig:NEC} as a function of inverse dimensionless spatial length $1/ma$ with finite temperature corrections.

Thermal effects decrease exponentially for small temperatures: for $m\beta\gg1$ we have 
\be
\sum_{n\in\mathbb{Z}}\frac{m^2a^2+8n^2\pi^2}{m^2a^3\omega_n}\left(\e^{\beta\omega_n/2}-1\right)^{-1}
\simeq \frac{\e^{-m\beta/2}}{am}~,
\ee
and in what follows we take the limit $\beta\to\infty$, in order to focus on NEC violating finite-length effects.  
It is then interesting to look at two asymptotic regimes for the length $a$:

\begin{itemize}
	\item[--] \underline{$ma\gg1$} In this case we have
    \be
	\frac{\rho+p}{ m^2}\simeq -\frac{\e^{-am}}{\sqrt{2\pi am}}
	-\frac{4}{3  }\sqrt\frac{am}{  \pi}~\e^{-2ma/3  }~,
	\ee
	and we can see that tunnelling effects are more important than the free-field Casimir effect; 
	\item[--] \underline{$ma\lesssim1$} In this case we have
	\be
	\label{eq:necTmal1}
	\frac{\rho+p}{m^2}\simeq-\frac{\pi}{3(am)^2}~,
	\ee   
	where tunnelling is negligible and the result is identical to the one obtained for a massless free field \cite{Bordag}.
\end{itemize}

\subsection{Entropy} \label{Sec:Entropy}

At zero temperature the classical thermal entropy vanishes, which can be seen with
\be
{\cal S}_\text{classical}=-\lim_{T\to0}\frac{\partial F_\text{true}}{\partial T}=0~.
\ee
There is a quantum contribution left though, which can be interpreted as the entropy for
the dilute gas of instantons/anti-instantons which relate the two vacua.
Taking into account the instanton fluctuation factors described above, 
the probability of having $n$ instanton/anti-instanton pairs may be read off from the partition function $Z$
\be\label{pn}
p_n=\frac{2}{Z}\frac{(\bar N)^{2n}}{(2n)!}\e^{-2W_\text{stat}(\beta/2)}
=\frac{1}{\cosh(\bar{N})}\frac{(\bar N)^{2n}}{(2n)!}~,
\ee
where $Z$, $\bar N$ and $W_\text{stat}$ are given by eqs. (\ref{Zsemi-classicalLowT3}), (\ref{Nbar}) and (\ref{Wstat}) respectively. The entropy of the dilute gas should be extensive and thus proportional to its number of degrees of freedom (although it is not proportional to the length $a$).
The entropy  for the systems of instantons and anti-instantons ${\cal S}_\text{tun}$ is then twice the entropy for the system of pairs, which is given by the usual sum over probabilities 
\be\label{entropy}
{\cal S}_\text{tun}=-2\sum_{n=0}^\infty p_n\ln(p_n)~.
\ee
In the limit $\bar N\gg1$, we find numerically that the entropy (\ref{entropy}) is asymptotically equivalent to 
\bea\label{quantumS}
{\cal S}_\text{tun}&\simeq& \ln(\bar{N})\\
&=& \ln(m\beta) -I_\text{inst}+\frac{1}{2}\ln\left(\frac{6}{\pi}I_\text{inst}\right)\nonumber~,
\eea
where $I_\text{inst}$ is the instanton action as given in \eqref{Instaction}, and the result matches the usual microcanonical entropy for $\bar N$ microstates.
As expected, one can also check that $\cal{S}_\text{tun}$ vanishes in the limit $a\to\infty$ (where $\bar{N}\to 0$, even for zero temperature),
since tunnelling is then completely suppressed. 
This behaviour is in correspondence with an ``effective third law of thermodynamics", 
where $1/a$ plays the role of a temperature. We will come back to this analogy later in this article.

For a finite length $a$, the entropy is non-zero with a logarithmic divergence in the zero temperature limit. Such logarithmic divergences in the zero temperature entropy of a quantum system are not new, such as the massless Casimir effect \cite{Cho:2022ngf} and it was argued in \cite{Brevik:2002gh} that such divergences should be removed.

The isothermal compressibility, defined as
\be \label{Compressibility}
K\equiv -\frac{1}{a}\frac{\partial a}{\partial p}
\equiv \frac{1}{a}\left( \frac{\partial^2F_\text{true}}{\partial a^2} \right)^{-1}~,
\ee 
is negative for all $a$, which is usually interpreted as a sign of instability. One may think that this instability is similar to the one obtained for a Van der Waals fluid experiencing an isothermal compression. 
If one assumes homogeneity of the Van der Waals fluid in a volume $V$, 
the bulk modulus $-V\partial p/\partial V$ is negative in a given range of volumes, which is not physical.
What happens is that the fluid separates into two phases, liquid and vapour, leading to the Maxwell construction, which corresponds to a constant-pressure plateau.
The position of this plateau is determined by identifying the chemical potentials in each phase. 
Also, a constant pressure leads to a vanishing compressibility, 
and the true free energy is convex, as expected for the Legendre transform of the internal energy. 
This constant-pressure plateau allows random regions of one vacuum or the other, in a proportion given by the volume, 
which varies between values corresponding to the first drop of liquid and the last bubble of vapour.

In our case, the system remains homogeneous though: 
the effective potential does not feature any plateau, but it has a unique minimum at $\left<\phi\right>=0$. 
An intuitive description is provided by weakly-interacting spins on a lattice, 
each with a random direction and a vanishing average value.\footnote{This is different from the high-temperature limit, where thermal fluctuations dominate over spin interactions and lead to a random spin distribution. The vanishing average spin discussed here happens at zero temperature, and is due to tunnelling instead}
A flat effective potential would be obtained in the limit $ma\to\infty$, where the tunnelling rate exponentially vanishes though, 
in which case one would have to wait an infinite amount of time for the true vacuum to settle.
As mentioned in section~\ref{sec:convexity}, in our situation the instability indicated by the negative compressibility would correspond to a spacetime expansion, 
if no environment was present to fix the length $a$.

\section{Black holes in 1+1 dilaton gravity}
\label{sec:dilatonicBHs}

The study of Hawking radiation including its backreaction on the spacetime geometry is an extremely difficult problem in $3+1$ dimensions. Motivated by dimensional reduction, it is possible to simplify the problem by studying the $1+1$ dimensional case. In particular, we focus on the classical  Callan, Giddings, Harvey and Strominger (CGHS) two-dimensional dilatonic black hole  model \cite{1992_Callan,1994_Fiola}. For the semi-classical description of the theory including backreaction, we consider the standard Polyakov term which represents the leading order quantum fluctuations in a $1/N$ expansion where $N$ is the number of matter fields. To find analytical solutions to the semiclassical theory it is necessary to introduce suitable counterterms in the action. To consider these counterterms, we introduce the one-parameter family of models that ranges between the Russo, Susskind and Thorlacius (RST) model \cite{1992_Russo} and the Bose, Parker, Peleg (BPP) model \cite{1995_Bose}, following the parameterization of the action presented in \cite{Cruz:1995zt}. We focus our study on the BPP model since it results in simpler expressions for the metric and the dilaton.

After describing the solutions of the semi-classical theory, this section delves into the implications for the stress-energy tensor and the entropy of two-dimensional black holes.

\subsection{Introduction to dilaton gravity}

In two dimensions, the Einstein-Hilbert action is just the Euler characteristic of the manifold (accordingly, $G_{\mu\nu}$  vanishes identically), and $1+1$ dimensional gravity is trivial. Since we want to capture aspects of the $3+1$ dimensional theory within a $1+1$ dimensional description, we use the dilaton field \cite{1984_Jackiw, 2002_Grumiller} which emerges from the compactification of higher dimensions. Here, we derive the dilaton from the dimensional reduction of spherically symmetric gravity in $3+1$ dimensions.

We consider the 3+1 Einstein-Hilbert action
\be \label{eqn:EH}
I^{(4)}_{EH}=\frac{1}{16\pi G^{(4)} } \int \text{d}^4 x\sqrt{-g^{(4)}} R^{(4)}\,, 
\ee
where $^{(4)}$ indicates the spacetime dimension and $G^{(4)}$ is Newton's constant. We consider the spherically symmetric ansatz
\be\label{eq:4dmetric}
\text{d}s_{(4)}^{2}=g_{ab}\text{d}x^a \text{d}x^b+\frac{e^{-2\phi(x^a)}}{\lambda^2}(\text{d}\theta^2+\sin^2\theta \text{d}\varphi^2)\,,
\ee
where the radius $r$ of the 2-sphere has been parametrized via a dilaton field $\phi(x^a)$, $r=\lambda^{-1}e^{-\phi}$. The parameter $\lambda$ is dimensionful and is introduced to get a dimensionless dilaton. Using this ansatz, we can write $R^{(4)}$ in terms of $R\equiv R^{(2)}$ and the four dimensional volume element in in terms of the two-dimensional volume element times the angular terms \cite{fabbri-navarro, 2022_Djordjevic}
\bea
R^{(4)}&=&R+2(\nabla\phi)^2+2\lambda^2e^{2\phi}-2e^{2\phi}\Box e^{-2\phi}\,,\nonumber\\
\text{d}^4x\sqrt{-g^{(4)}}&=&\text{d}^2x\,\text{d}\theta\, \text{d} \varphi \sqrt{-g}\frac{e^{-2\phi}}{\lambda^2}\sin^2\theta\,.
\eea
Integrating out the angular part, the resulting dilaton action is \cite{fabbri-navarro, 2022_Djordjevic}
\be
\label{eq:Sdil}
I_{D}=\frac{1}{4\pi\lambda^2 G^{(4)}}\int \text{d}^2x\sqrt{-g}\left(e^{-2\phi}\left(R+2(\nabla\phi)^2\right)+2\lambda^2\right)~,
\ee
where we see that $\lambda$ plays the role of a cosmological constant in the reduced theory. We define a two dimensional Newton's constant $G^{(2)}=\lambda^2 G^{(4)}$ and work in units where $G^{(4)}=\frac{1}{2\lambda^2}$.

To simplify the theory in such a way that it is possible to find an exact analytical solution, we work with the action\footnote{$I_{\phi}$ can be exactly derived by integrating out the angular part of near-extreme, magnetically charged black holes in four-dimensional dilaton gravity \cite{1992_Giddings,1992_Callan}.} 
\be
\label{eq:phi}
I_{\phi}=\frac{1}{2{\pi}}\int \text{d}^2x\sqrt{-g}e^{-2\phi}\left(R+4(\nabla\phi)^2+4\lambda^2\right)\,,
\ee
where we have modified the potential term of the dilation and the coefficient of the kinetic term as compared with (\ref{eq:Sdil}). Despite the changes, this dilaton theory still has black holes and Hawking radiation \cite{1992_Strominger,1992_Callan,1994_Fiola}. We work in conformal gauge, 
\be \label{eq:conformalgauge0}
\text{d}s^2=-e^{2\eta\,}\text{d}x^{+}\text{d}x^{-}\,,
\ee
with null coordinates $x^{\pm}$. The EoM resulting from the variation of the action with respect to $\eta$ and  $\phi$ can be conveniently written in terms of $2(\eta-\phi)$ and $e^{-2\phi}$, namely
\bea
\label{eq:TMmclass}\partial_{+} \partial_{-} e^{-2\phi}+\lambda^2 e^{2( \eta-\phi)}&=&0\,, \\
\label{eq:phiclass}2e^{-2\phi} \partial_+ \partial_{-}(\eta-\phi)+\partial_+\partial_- e^{-2\phi}+\lambda^2 e^{2(\eta-\phi)}&=&0\,. 
\eea
Additionally, we derive the following constraints from  the variation of the action with respect to the $(\pm, \pm)$ components of the metric $g_{\mu \nu}$ 
\be
\partial_{\pm}^2 e^{-2\phi}-2\partial_{\pm}(\eta-\phi)\partial_{\pm}e^{-2\phi}=0\,.
\ee
Combining equations \eqref{eq:phiclass} and \eqref{eq:TMmclass} we get the free field equation
\be
\label{eq:ffeq}
2\partial_+ \partial_-(\eta-\phi)=0\, ,
\ee
which has solutions of the form $2(\eta-\phi)=h(x^+)+s(x^-)$. In the Kruskal gauge,  the remaining freedom is fixed by making $h(x^+)=s(x^-)=0$, i.e. $\eta=\phi$. This model has black hole solutions which, in the Kruskal gauge, are (see \cite{1992_Callan, fabbri-navarro} for a detailed discussion)
\bea \label{eq:dilatonicBHKruscal}
ds^2&=&-\frac{dx^+dx^-}{(M/\lambda-\lambda^2 x^+x^-)}\, ,\\
\label{eq:dilatonKR}\eta &=&\phi=-\frac{1}{2}\ln(\frac{M}{\lambda}-\lambda^2 x^{+}x^{-})\,,
\eea
where $M$ is an integration constant that corresponds to the ADM mass of the black hole \cite{1992_Callan}. This metric has a curvature singularity at $\lambda^2x^+x^-=M/\lambda$ and horizons at $\lambda^2x^+x^-=0$, see figure \ref{fig:BHdia}.  The surface gravity can be easily computed, and reads $\kappa=\lambda$. Therefore, the black hole temperature is  \be \label{eqn:lambda}
T=\frac{\lambda}{2\pi}\, .
\ee
In contrast to the black hole temperature in four dimensions, the two-dimensional black hole temperature is independent of the mass $M$.  We use these results to evaluate the NEC and the entropy.

\begin{figure}[h]
	\centering
	\begin{tikzpicture}[scale=0.5]
		\node (I)    at ( 4,0)   {I};
		\node (II)   at (-4,0)   {II};
		\node (III)  at (0, 2.5) {III};
		\node (IV)   at (0,-2.5) {IV};
		
		\path  % Four corners of left diamond
		(II) +(90:4)  coordinate[label=90:$i^+$]  (IItop)
		+(-90:4) coordinate[label=-90:$i^-$] (IIbot)
		+(0:4)   coordinate                  (IIright)
		+(180:4) coordinate[label=180:$i^0$] (IIleft)
		;
		\draw (IIleft) -- 
		node[midway, above left]    {$\cal{J}^+$}
		node[midway, below, sloped] {}
		(IItop) --
		node[midway, above, sloped] {}
		(IIright) -- 
		node[midway, below, sloped] {}
		(IIbot) --
		node[midway, above, sloped] {}
		node[midway, below left]    {$\cal{J}^-$}    
		(IIleft) -- cycle;
		
		\path % Four conners of the right diamond (no labels this time)
		(I) +(90:4)  coordinate[label=90:$i^+$] (Itop)
		+(-90:4) coordinate[label=-90:$i^-$] (Ibot)
		+(180:4) coordinate (Ileft)
		+(0:4)   coordinate[label=0:$i^0$] (Iright)
		;
		% No text this time in the next diagram
		\draw  (Ileft) --node[midway, above, sloped] {$x^-=0$} (Itop) -- node[midway, above right]    {$\cal{J}^+$}
		node[midway, below, sloped] {}(Iright) -- node[midway, above, sloped] {}
		node[midway, below right]    {$\cal{J}^-$}(Ibot) --node[midway, above, sloped] {$x^+=0$} (Ileft) -- cycle;
		
		% Squiggly lines
		\draw[decorate,decoration=zigzag] (IItop) -- (Itop)
		node[midway, above, inner sep=2mm] {};
		
		\draw[decorate,decoration=zigzag] (IIbot) -- (Ibot)
		node[midway, below, inner sep=2mm] {};
	\end{tikzpicture}
	\caption{Penrose diagram for a static two-dimensional dilatonic black hole.}
	\label{fig:BHdia}
\end{figure}

\subsection{Adding quantum matter: static black holes}

Now we add $N$ massless scalar fields $f_i$ and we have the total action 
\be
\label{eq:S0}
I_{0}=I_{\phi}+I_f=\frac{1}{2\pi}\int \text{d}^2x\sqrt{-g}\left[e^{-2\phi}\left(R+4(\nabla\phi)^2+4\lambda^2\right)
-\frac{1}{2}\sum_{i=1}^N(\nabla f_i)^2\right]\,.
\ee
This action corresponds to the CGHS model. We continue the analysis in conformal gauge \eqref{eq:conformalgauge0}. In these coordinates, the classical stress-energy tensor for the fields $f_i$ is 
\be
T_{\pm\pm}= \frac{1}{2}\sum_{i=0}^N(\partial_{\pm}f_i)^2 \,.
\ee
The next step is to quantise the theory. We want to focus on static solutions, $\langle f_i\rangle=0$, and look at the one-loop quantum corrections to the stress-energy tensor in different vacuum states. The quantum corrections of the different fields that we have seen in the CGHS model contribute to the semiclassical theory. In order to make the analysis feasible, we assume that the number of matter fields $N$ is very large and calculate the effective action at leading order in an expansion in $1/N$. In this limit, the quantum fluctuations of the dilaton and the metric can be ignored and we only have to consider the one-loop correction of the matter fields to the stress-energy tensor \cite{1992_Harvey, 1994_Fiola,Mikovic:1997mz}. This one-loop correction to $T_{\mu\nu}$ due to the $N$ massless scalar fields can be evaluated using the trace anomaly, which relates the expectation value of the stress-energy tensor and the Ricci scalar \cite{1977_Christensen}
\be \label{eq:anonaly00}
\langle {T}\rangle= \frac{N}{24}R\,.
\ee
In conformal gauge, the trace anomaly leads to
\be
\label{eq:Tpm}
\langle {T}_{+-}\rangle =-\frac{N}{12}\partial_+\partial_-\eta\,.
\ee
In addition, we can use $\langle {T}_{+-}\rangle$ in \eqref{eq:Tpm} together with the conservation of the stress-energy tensor to determine $\langle {T}_{\pm\pm}\rangle$
\be \label{eq:Tpp}
\langle {T}_{\pm\pm}\rangle=-\frac{N}{12} \, \left(\partial_{\pm}\eta \partial_{\pm}\eta-\partial^2_{\pm}\eta + t_{\pm}\right)\, ,
\ee
where $t_{\pm}$ is fixed by boundary conditions (vacuum choice). We will analyse two vacuum choices: the Hartle-Hawking vacuum, which describes thermal equilibrium at infinity and is given by $t_{\pm}=0$, and the Boulware vacuum, which describes empty space at infinity and is given by the boundary conditions $t_{\pm}=-\frac{1}{4(x^{\pm})^2}$.

Alternatively, the expectation value of the stress-energy tensor can be obtained by functional differentiation of an effective action, the Polyakov action 
\be
\label{eq:TdSdG}
\langle T_{\mu \nu}\rangle= -\frac{2\pi}  {\sqrt{-g}} \frac{\delta I_P}{\delta g^{ \mu \nu}}\, ,
\ee
where
\be
\label{eq:Spol1}
I_P=-\frac{N}{96\pi}\int \textrm{d}^2 x \sqrt{-g(x)} \int \textrm{d}^2 y \sqrt{-g(y)} \, R(x) G(x,y) R(y)\,.
\ee
$G(x,y)$ is the Green's function for the differential operator $\Box_g$. $I_P$ incorporates the backreaction of the quantum fluctuations of the matter fields on the metric. By writing $I_P$ in the conformal gauge, we can derive immediately Eqs. \eqref{eq:Tpm} and \eqref{eq:Tpp} from \eqref{eq:TdSdG}. We will use these expressions later on in the evaluation of the NEC.

It is convenient to convert the non-local Polyakov action into a local one by introducing an auxiliary scalar field $\varphi$ constrained to obey the equation $\Box_g \varphi =-R$ (see for example eq.~(5.56) in \cite{fabbri-navarro}). By doing this, the local action is
\be
I_P=-\frac{N}{96\pi}\int \textrm{d}^2 x \sqrt{-g(x)}\left[(\nabla \varphi)^2-2R\varphi\right]\,.
\ee
In conformal gauge, the equation of motion for $\varphi$ has the following solution
\be
\label{eq:varFIsol}
\varphi(x^{\pm})=2\eta(x^{\pm}) -2\left(\varphi_+(x^{+})+\varphi_-(x^{-})\right)\,,
\ee
where $\varphi_+(x^{+})$ and $\varphi_-(x^{-})$ are solutions of
\be
\label{eq:auxFI}
\partial_{\pm}^2\varphi_{\pm}-(\partial_{\pm}\varphi_{\pm})^2=t_{\pm}(x^{\pm})\,.
\ee

We are also interested in the entropy of the system. As we showed in \eqref{eq:dilatonicBHKruscal}, this theory has black hole solutions. Therefore, the total entropy of the system consists of two terms: the geometric black hole entropy, which is the $1+1$ dimensional equivalent of the Bekenstein-Hawking entropy, and the von Neuman entropy, associated to the quantum fields outside the horizon, usually called fine-grained entropy \cite{1994_Fiola, 1994_Myers, 2023_Hirano}. 
For general diffeomorphism invariant theories, it is possible to evaluate both of these entropies using the method developed in  ~\cite{1994_Iyer} and particularised to the context of two-dimensional gravity in ~\cite{1994_Myers, 2022_Yu, 2023_Hirano}. In this way, both entropies can be evaluated in a geometrical way by calculating the derivatives of the Lagrangian associated with the different contributions to the action with respect to the curvature
\bea
\label{eq:SBH}
\mathcal{S}_\phi&=& \left.\frac{4\pi }{\sqrt{-g}}\frac{\partial \mathcal{L}_{\phi}}{\partial R}\,\right|_H=2 e^{-2\phi} \,\Big|_H\\
\label{eq:Smatter}
\mathcal{S}_P &=& \frac{4\pi}{\sqrt{-g}}\frac{\partial \mathcal{L}_P}{\partial R}\Bigg|_H=\frac{N}{12} \, \varphi\,\Big|_H\, ,
\eea
where $|_H$ means that these quantities should be evaluated at the horizon. It can be shown that $\mathcal{S}_\phi$ and $\mathcal{S}_P$ are equivalent to the black hole and the fine-grained entropy respectively  \cite{1994_Fiola, 2021_Pedraza, 2023_Hirano}. Equivalently, the entropy can be evaluated using the Euclidean path integral approach, as done in  \cite{Hayward:1994dw} for the RST model.

In what follows, we proceed to evaluate the NEC and the entropy. We start with the simpler case without backreaction, and then we study how backreaction modifies the results.

\subsection{Without backreaction, the CGHS model}
For the case without backreaction, the  background spacetime is described by \eqref{eq:S0}, i.e. the CGHS model. Since we are interested in static black holes, we focus on the case where $\langle f\rangle=0$. As we  have seen, under these conditions, the solution is an eternal black hole of mass $M$.

\subsubsection{Null energy condition}\label{sec:NECclas}

The null energy condition \be \langle  T_{\mu \nu}\rangle \ell^\mu \ell^\nu \geq 0\, ,\ee gives two equations that, in the conformal gauge, are proportional to the diagonal components of the stress-energy tensor $\langle  T_{\pm\pm}\rangle$ . In what follows, we will focus on the exterior region, so we assume $x^+>0$ and $x^{-}<0$. For the Hartle-Hawking vacuum ($t_{\pm}=0$), using the dilatonic metric \eqref{eq:dilatonKR} in  eq. \eqref{eq:Tpp}, we find\footnote{The condition $\langle  T_{\mu \nu}\rangle \ell^\mu \ell^\nu \geq 0$ is defined up to a positive overall factor. In this case, we find it useful to compute $(\lambda x^{\pm})^2\langle   T_{\pm\pm}\rangle$ instead of $\langle   T_{\pm\pm}\rangle$, since it is the quantity related to an asymptotic observer at infinity. This quantity comes directly from the transformation law for tensors $T_{vv}=(dx^{+}/dv)^2 \,T_{++}$ (and similarly for the other components).}
\be \label{eq:NECHH0}
NEC^{\pm}_H=(\lambda x^{\pm})^2\langle H|  T_{\pm\pm}|H\rangle=
\frac{N\, \lambda^2}{48}\, ( \lambda x^{\pm})^2( \lambda x^{\mp})^2\,e^{4\eta} >0\, ,
\ee
while for the Boulware vacuum we obtain 
\be
\label{eq:TsAH}
NEC^{\pm}_B=(\lambda x^{\pm})^2\langle B|  T_{\pm\pm}|B\rangle=-\frac{N\, \lambda^2}{48}\left(1-( \lambda x^{\pm})^2 ( \lambda x^{\mp})^2\,e^{4\eta}\right)
<0\, .
\ee
We can easily check that the difference between the Hartle–Hawking and Boulware stress energy tensors is just a thermal distribution of massless particles at the Hawking temperature\footnote{See \cite{Visser:1996ix} for a similar analysis in a Schwarzschild background.} 
\be \label{eq:NECthermal}
NEC_H^{\pm}-NEC_B^{\pm}=N\frac{\lambda^2}{48}=N\frac{\pi^2}{12}\, T^2\,,
\ee
where $T=\lambda/2\pi$ is the black hole temperature as given in eq. \eqref{eqn:lambda}.
The extra $N$ factor appears because we are considering $N$ conformal fields. For convenience, we can write the stress-energy tensor in the Hartle-Hawking vacuum as
\be \label{eq:NECHH01}
NEC_H^{\pm}= \frac{N \lambda^2}{48}-\frac{N  \lambda^2}{48}\left(1-( \lambda x^{\pm})^2 ( \lambda x^{\mp})^2\,e^{4\eta}\right) >0\,.
\ee
In this expression, we see that there is a negative contribution coming from pure vacuum effects plus a positive contribution coming from thermal effects.
As a final note, we point out that  the quantity $\left(1-( \lambda x^{\pm})^2 ( \lambda x^{\mp})^2\,e^{4\eta}\right)$ vanishes as $x^{\pm}\to \pm\infty$ and tends to 1 as $x^{\pm}\to 0$. It means that  $(\lambda x^{+})^2\langle H|  T_{++}|H\rangle\to N \lambda^2/48$ asymptotically (constant thermal flux), and goes to zero at the horizon (thermal bath in thermal equilibrium with the black hole, so the fluxes cancel). We can remove the contribution from the thermal bath to compute the NEC vacuum contribution of the black hole
\be \label{rho+pBHquantum}
NEC_{\text{BH}}=(\lambda x^{\pm})^2\langle H|  T_{\pm\pm}|H\rangle -\left(\frac{N \, \lambda^2}{48}\right)=-\frac{N \, \lambda^2}{48}\left(1-( \lambda x^{\pm})^2 ( \lambda x^{\mp})^2\,e^{4\eta}\right) <0\, ,
\ee
which results in a negative contribution to the NEC due to vacuum effects. On the horizon
\be \label{NECQuantumBH}
NEC_{\text{BH}} = -\left(\frac{N \,{\pi^2} T^2}{12}\right) \,.
\ee

\begin{figure}[h]
	\centering
	\includegraphics[width=10cm]{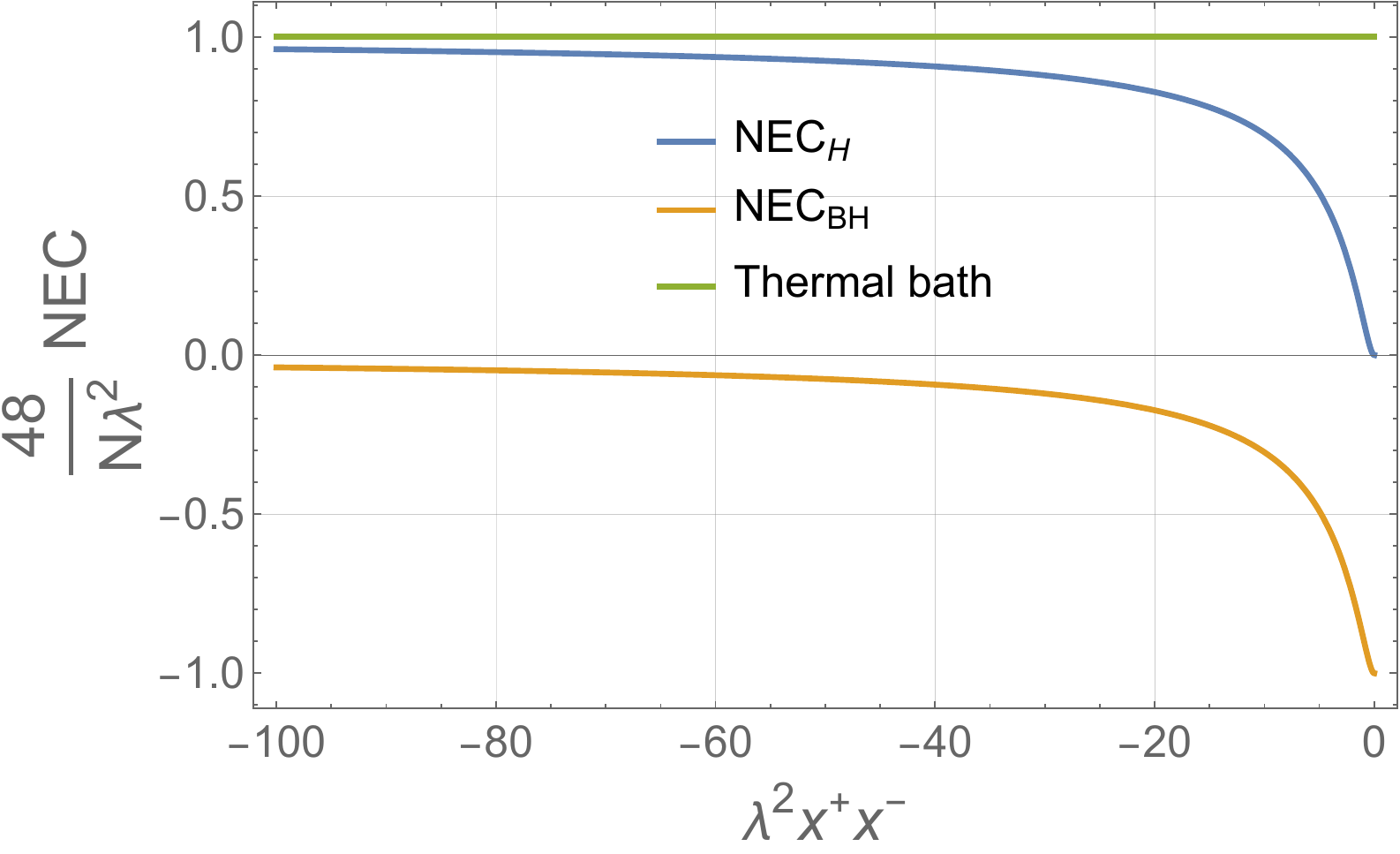}
	\caption{ \small{We plot the (normalized) energy condition $\frac{48}{N \lambda^2 }$(NEC) as a function of $\lambda^2 x^+\, x^-$ for  the BPP and classical models. The green curve corresponds to the  thermal bath eq. \eqref{eq:NECthermal}, the orange curve corresponds to the contribution from  vacuum effects eq. \eqref{eq:TsAH},  and the total NEC  eq. \eqref{eq:NECHH01} is represented in blue. The black hole mass is $M/\lambda=2$.}}
	\label{fig:NECBH}
\end{figure}

In figure~\ref{fig:NECBH} we summarise the results for the NEC.

\subsubsection{Entropy and thermodynamics}\label{sec:Sclas}

Inserting the dilaton solution \eqref{eq:dilatonKR} into \eqref{eq:SBH}, the black hole entropy is 
\be
\label{eq:ActionBH}
\mathcal{S}_{BH}=\mathcal{S}_{\phi}=\frac{2 M}{\lambda}=\frac{M}{\pi T}\, .
\ee
From the dimensional reduction \eqref{eq:4dmetric} relating the radius and the dilaton, we can interpret this term as $1/4$ of the horizon area of the classical $3+1$-dimensional black hole. Remember that in the dimensional reduction \eqref{eq:4dmetric}, $r=\lambda^{-1} e^{-\phi}$ and $\mathcal{S}_{\phi}=2 e^{-2\phi_H}=\frac{2\lambda^2}{\pi}\,\frac{4 \pi r_H^2}{4}$ (see eq. \eqref{eq:SBH}).

To evaluate the entropy associated with the $N$ massless fields surrounding the black hole we use \eqref{eq:Smatter}. Although we are not considering backreaction in this subsection, we can still evaluate $\mathcal{S}_P$ using  the classical solution \eqref{eq:dilatonKR} to obtain the fine-grained entropy of the matter fields. In the Hartle-Hawking vacuum ($t_{\pm}(x^{\pm})=0$),  the most general solution for the auxiliary field $\varphi$ is [see Eqs. \eqref{eq:varFIsol} and \eqref{eq:auxFI}]  
\be 
\label{eq:auxiliarfield}
\varphi=2\eta+2\ln[(a^+ x^+ + b^+)(a^- x^- + b^-)]~,
\ee
where $a^+,a^-,b^+,b^-$ are constants of integration. 

As discussed after \eqref{eq:Smatter}, $\mathcal{S}_P$ %accounts for 
can be related to the von-Neuman entropy associated to the quantum fields outside the horizon, known as fine-grained entropy $\mathcal{S}_{FG}$. The fine-grained entropy for the RST was calculated, for example, in \cite{1994_Fiola} [see eq. (93)]
\be
\mathcal{S}_{FG}=\left.\frac{N}{6}\eta\right|_H +\frac{N}{12}\ln(\frac{-x_{max}^{+}x_{max}^{-}}{\delta^2})\,,
\ee
where $x_{max}^+$ and $x_{max}^-$ are infrared cutoffs in Kruskal coordinates introduced to regulate the contribution of long-wavelength modes to the entropy, and $\delta$ is a short distance cut-off introduced to regularise the (ultraviolet) logarithmic divergence. We have freedom to choose the constants $a^{\pm}$ and $b^{\pm}$ in \eqref{eq:auxiliarfield} in such a way that, when replaced in \eqref{eq:Smatter} and evaluated at the horizon,
\be \mathcal{S}_P=\mathcal{S}_{FG}\, \ee 
(see for example \cite{2021_Pedraza}). Following \cite{1994_Fiola}, we can see that the logarithmic term in $\mathcal{S}_{FG}$ can be related to the entropy of a thermal bath.

 Indeed, if we rewrite the logarithmic term in light-cone coordinates 
\be
x^+=\lambda^{-1} e^{\lambda \sigma^+}\,, \quad \text{and}\quad x^-=-\lambda^{-1} e^{-\lambda \sigma^-}\,,
\ee
we obtain
\be \label{eq:Sthermal0}
\frac{N}{12}\ln(\frac{-x_{max}^{+}x_{max}^{-}}{\delta^2})=\frac{N}{12}  \lambda (\sigma^+-\sigma^-)_{max}=\frac{N}{6} \lambda \sigma_{max}=2\frac{N}{6} T L=\mathcal{S}_{thermal}\,,
\ee
where we interpret $\sigma_{max}$ as the length of a box of size $L$.

This is exactly the entropy of a thermal bath in a one dimensional box of length $2L$, as seen by an asymptotic observer.

This result allows us to rewrite the entropy of the Polyakov term as 
\be
\mathcal{S}_P=\mathcal{S}_{quantum} + \mathcal{S}_{thermal}\, ,
\ee
where 
\be
\label{eq:Smatterclas}
\mathcal{S}_{quantum}=\frac{N}{6} \eta|_H=-\frac{N}{12}\ln\left(\frac{M}{\lambda}\right)\,.
\ee

In what follows, we will subtract the thermal contribution to the entropy, since we want to focus on vacuum effects. Now, the total entropy of the system is (excluding $\mathcal{S}_{thermal}$)
\be
\mathcal{S}_{tot}=\mathcal{S}_{BH}+\mathcal{S}_{quantum}=\frac{2 M}{\lambda}-\frac{N}{12}\ln\left(\frac{M}{\lambda}\right)\, ,
\ee
where we have used that, in the Kruskal gauge, $\phi=\eta$ is given by eq. \eqref{eq:dilatonKR}.

\subsection{With backreaction, the BPP model}

As we discussed, the backreaction on the metric of the quantum fluctuations of the matter field can be considered by adding $I_P$ to the action $I_{\phi}$. However, the EoM from $I_{\phi}+I_P$ can not be solved analytically. To address this problem, we can add the following extra term to the action \cite{Cruz:1995zt}
\bea \label{eq:Iextra}
I_{extra}&=&\frac{N}{24 \pi} \int \textrm{d}^2 x \sqrt{-g}\left[(1-2 b)(\nabla \phi)^2+(b-1) \phi R\right] \nonumber\\
&=&\frac{N}{24 \pi} \int \textrm{d}^2 x \left(-2(1-2b)\partial_+\phi \partial_-\phi+4(b-1)\phi\partial_+\partial_-\eta\right)\,. 
\eea
This results in a family of models characterized by the parameter $b$. $I_{extra}$ being local modifies the local dynamics but not the global properties. For $b=1/2$ we recover the RST model \cite{1992_Russo} and for $b=0$ we recover the BPP model \cite{1995_Bose}. By introducing the Liouville fields \cite{Cruz:1995zt}
\bea
\Omega &=&\sqrt{\frac{N}{12}}\, b\phi+\sqrt{\frac{12}{N}}e^{-2 \phi}, \label{eq:change1}\\
\chi&=&\sqrt{\frac{N}{12}} \, \eta+\sqrt{\frac{N}{12}}(b-1)\phi+\sqrt{\frac{12}{N}}e^{-2 \phi}\,,\label{eq:change2}
\eea
we can rewrite our family of models as a Liouville theory that can be solved analytically. The EoM for the action $I_{\phi}+I_{P}+I_{extra}$ in terms of the Liouville variables are 
\bea
 \partial_{+} \partial_{-} \chi&=&-\lambda^2\sqrt{\frac{12}{N}}\,  e^{\sqrt{\frac{48}{N}}\,(\chi-\Omega)}, \\
 \partial_{+} \partial_{-} \Omega&=&-\lambda^2\sqrt{\frac{12}{N}}\, e^{\sqrt{\frac{48}{N}}\,(\chi-\Omega)},
\eea
which implies 
\be
\partial_+\partial_-(\chi-\Omega)=0\, .
\ee
The constraint equations become
\be
-\partial_{\pm}\chi \partial_{\pm}\chi+\sqrt{\frac{N}{12}} \partial_{\pm}^2\chi+\partial_{\pm} \Omega \partial_{\pm} \Omega -\frac{N}{12}t_{\pm}=0\, .
\ee
In the Kruskal gauge $\Omega=\chi$ and for the Hartle-Hawking vacuum ($t_{\pm}=0$) we have the solution
\be
\label{eq:Om}
\sqrt{\frac{N}{12}}\,\Omega=\frac{M}{\lambda }-\lambda^2x^{+} x^{-}\, .
\ee
From now on, we focus on the model $b=0$, i.e. the BPP model, since it results in a simpler solution. Taking $b=0$ and \eqref{eq:Om} into the Liouville variables \eqref{eq:change1} and \eqref{eq:change2}, it is immediate to find
\be
\phi=\eta=-\frac{1}{2}\ln\left(\frac{M}{\lambda}-\lambda^2 x^+x^-\right)\,.
\ee
These solutions are the same as the classical solutions \eqref{eq:dilatonKR}, which indicates that the metric and the dilaton do not have quantum corrections in the BPP model.

\subsubsection{Null energy condition}
The results for the NEC in section \ref{sec:NECclas} don't change when we consider the BPP model in the Hartle-Hawking vacuum since the metric is not affected by the quantum corrections. Figure \ref{fig:NECBH} summarizes the results for the NEC in the BPP model and the classical results.

\subsubsection{Entropy and thermodynamics}
\label{sub:bhentropy}

The evaluation of the entropy follows the discussion in section \ref{sec:Sclas}. Since we included the $I_{bpp}=I_{extra}(b=0)$ term in the action, we have an additional contribution to the entropy 
\be
\mathcal{S}_{bpp}=\frac{4\pi}{\sqrt{-g}}\frac{\partial \mathcal{L}_{bpp}}{\partial R}\Bigg|_H=  -\frac{N}{6}\phi\Bigg|_H=\frac{N}{12}\ln\left(\frac{M}{\lambda}\right)\,.
\ee
The total entropy of the semi-classical system is (as before, we omit the $\mathcal{S}_{thermal}$ contribution of $\mathcal{S}_P$)
\be
\label{eq:Stot}
\mathcal{S}_{tot}=\mathcal{S}_\phi+\mathcal{S}_{bpp}+\mathcal{S}_{quantum}=2 e^{-2\phi} \Big|_H=\frac{2 M}{\lambda}\, ,
\ee
where we have used that, in the Kruskal gauge $\phi=\eta$, and the dilaton is given by \eqref{eq:dilatonKR}. Note that $\mathcal{S}_{bpp}$ cancels with $\mathcal{S}_{quantum}$. We find that  $\mathcal{S}_{tot}$ is equal to the $\mathcal{S}_{BH}$ for the case without backreaction \eqref{eq:ActionBH}. It means that the total entropy of the semi-classical system is exactly the entropy of a dilatonic black hole of mass $M$ at a temperature $\lambda/2\pi$. This is not surprising since the solution to the semi-classical equations in the Hartle-Hawking vacuum state is precisely a black hole of mass $M$ at a temperature $\lambda/2\pi$. However, we can still split the total entropy into two parts: the entropy of the black hole, and the entropy of the matter sector
\bea\label{eq:Sbhsc}
\mathcal{S}_{BH}&=& \mathcal{S}_{\phi}+ \mathcal{S}_{bpp}=\frac{2 M}{\lambda}+\frac{N}{12}\ln\left(\frac{M}{\lambda}\right)\, ,\qquad \\
\label{eq:Smatsc}\mathcal{S}_{quantum}&=& -\frac{N}{12} \ln \frac{M}{\lambda}\, .
\eea
The matter entropy $\mathcal{S}_{quantum}$ is negative just because we removed the log divergent contribution coming from the thermal bath. Let us note that $\mathcal{S}_{BH}$ has a quantum correction as compared to the black hole entropy without backreaction \eqref{eq:ActionBH} coming from $\mathcal{S}_{bpp}$.

We finally comment on a related quantity, which is the black hole heat capacitance for the complete system  
(see Eqs. \eqref{eq:Stot} and \eqref{eq:Sthermal0})
\be
C_{tot}= T\frac{\partial (\mathcal{S}_{tot}+\mathcal{S}_{thermal})}{\partial T}=\frac{\pi N T L}{3}-\frac{M}{\pi T}~,
\ee
and which is consistently positive, due to the dominant contribution of the thermal bath. Thus the full system is thermodynamically stable.

\section{Comparison}
\label{sec:comparison}

In sections \ref{sec:Euclidean} and \ref{sec:dilatonicBHs} we discussed the thermodynamics of finite volume effects in QFT and dilatonic black holes respectively. These effects seem at first glance unrelated. However, in both systems we have NEC violation and finite quantum  entropy. But the analogy runs even deeper; we observe that the effect of the temperature $T$ in black holes is similar to the effect of the inverse length scale $1/a$ in tunneling. To see this, we compare two physical quantities, which are the NEC and the rate of change in entropy, when the environment is modified.\\

\noindent {\it NEC violation.}\\ 
For tunneling, the NEC was computed in section~\ref{Sec:NEC}. We will compare in the regime where $ma \lesssim 1$, where the NEC becomes
\be
NEC_{\text{tun}} \simeq-\frac{\pi}{3a^2}~.
\ee   
It is interesting to note that in this regime the Casimir effect is dominant over the tunneling effects. 
The result on the black hole horizon for any ratio $T/M$ in the BPP model is computed in section~\ref{sec:dilatonicBHs} 
\be
NEC_{\text{BH}} = -\frac{N \,\pi^2 T^2}{12}~.
\ee
We notice the clear equivalence of the dominant terms in the regime of interest: NEC violation is proportional to $1/a^2$ and $T^2$ for the tunneling and the black hole respectively. Physically, this is partly a manifestation of the effect that the NEC violation is due to finite volume effects in one case and the non-zero temperature of the black hole in the other.\\

\noindent {\it Entropy}\\
For the tunneling case we found (Sec.~\ref{Sec:Entropy}) 
\be
\label{eq:Stunf}
\mathcal{S}_{tun}=\ln(m\beta)-\frac{2ma}{3}+\frac{1}{2}\ln\left(\frac{4ma}{\pi}\right)~.
\ee
The black hole entropy is (Sec.~\ref{sub:bhentropy})
\be
\label{eq:Sbhf}
\mathcal{S}_\text{BH}+\mathcal{S}_{thermal}=\frac{N\pi}{6}T (2L) +\frac{M}{\pi T}+\frac{N}{12}\ln\left(\frac{M}{2\pi T}\right)~.
\ee
We note that both entropies have a term that diverges in the limit of zero temperature for the tunneling and the limit of infinite volume for the black hole. These terms are $\ln(m\beta)$ for tunneling and $\frac{N\pi}{6}T (2L)$ for the black hole. We will remove these terms as we want to compare the quantum entropies of the two systems without infinite contributions. Then the entropy in both cases can be negative.

Instead of comparing the entropies as they are, we will focus on the rate of change of the entropy in terms of the relevant parameter $w$ for each system
\be
r \equiv w\frac{\partial \mathcal{S}}{\partial w} \,.
\ee

So we consider the rate of change in entropy when the length $a$ varies
\be\label{Lengthrate}
r_\text{tun}\equiv a\frac{\partial\mathcal{S}_\text{tun}}{\partial a}=\frac{1}{2}-\frac{2m}{3}a~.
\ee
For the black hole case, the rate of change is in terms of the temperature
\be\label{Temperaturerate}
r_{BH}\equiv T\frac{\partial\mathcal{S}_\text{BH}}{\partial T}=-\frac{N}{12}-\frac{M}{\pi T}~.
\ee
As for NEC violation, we observe that there is a clear equivalence of the rates of change in entropy when one maps $a$ to $1/T$.
This is not an effect that can be explained by simple dimensional analysis: the quantities $ma$ and $T/M$ are dimensionless and thus any combination could appear in these expressions.

\begin{figure}
    \centering \includegraphics[width=0.5\linewidth]{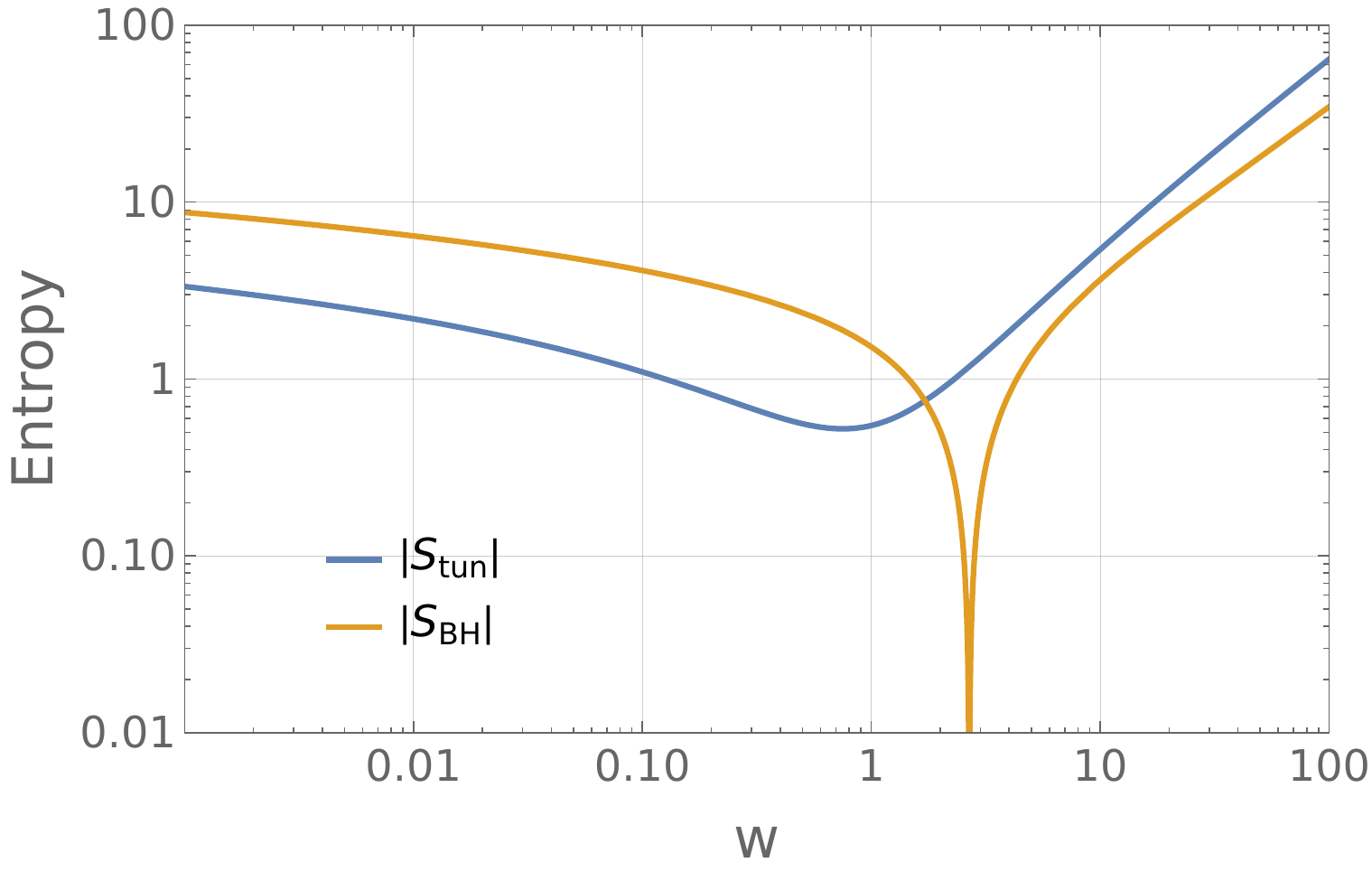}
    \caption{Tunneling \eqref{eq:Stunf} and black hole \eqref{eq:Sbhf} entropies as a function of $\text{w}=ma$ and $\text{w}=\frac{M}{T}$, respectively. We have removed the $\ln(m\beta)$ divergence and set $N=12$.}
    \label{figA}
\end{figure}

Fig. \ref{figA} shows the absolute value of $\mathcal{S}_{tun}$ and $\mathcal{S}_{BH}$ as a function of $ma$ and $M/T$, respectively. Without the term $\ln(m\beta)$, $\mathcal{S}_{tun}$ is always negative because of the dominant contribution of the negative linear term. For $\mathcal{S}_{BH}$, the positive linear term becomes dominant when $M/T$ exceeds $\frac{N\pi}{12}W_0(\frac{24}{N})$. The divergence observed in the logarithmic plot indicates a change in the sign of $\mathcal{S}_{BH}$ at higher temperatures, where the logarithmic term takes over. Although these quantities can be negative, the total entropy remains positive due to the contributions of $\ln(m\beta)$ and $S_{thermal}$.

We should briefly comment on a relevant comparison between the Casimir effect and black hole thermodynamics in 
\cite{Widom:1998pn,Sassaroli:1998ug}. There, the authors used isothermal compressibility instead of the rate of change of the entropy but also found an analogy.

\section{In higher dimensions}
\label{sec:higherd}

While the results in this article are for $1+1$-dimensions we believe this analogy can be extended to dimension $d+1$. 
Here, we present some results that point to a correspondence and comment on future work.

\subsection{Finite size effects}

This section is based on some results derived in \cite{Ai:2024taz} and \cite{Alexandre:2024gat}, 
that we generalise here to $d$-dimensions, in order to support the extension of our results to higher dimensions.

\subsubsection{Finite size effects in a higher dimensional torus}

In higher space dimensions, the potential of the one-dimensional action \eqref{Sbare} should be upgraded to
    \be 
    U(\phi)=\frac{\lambda}{24}(\phi^2-v^2)^2~,
    \ee
and the mass is now expressed in terms of the quartic self-interaction coupling constant $\lambda$ and the vacuum expectation value $v$ via the relation
    \be 
    m^2=\frac{\lambda v^2}{3}~.
    \ee
The instanton action in $(d+1)$-dimensions then becomes
    \be\label{Id+1}
    I_\text{inst}=\frac{2m^3V}{\lambda}~, 
    \ee
where $V$ is the $d$-dimensional spatial volume and the mass dimension of the coupling constant is $[\lambda]=3-d$.

For a $d$-dimensional torus and a massive scalar field, tunnelling is suppressed exponentially with the volume, 
whereas the Casimir effect is suppressed exponentially with a typical length of the confining space. 
As a consequence, similar order of magnitude for these effects is expected for $d=1$ only, as we have seen in previous sections. 
In the situation of a torus with identical sizes $a$, it was shown in \cite{Ai:2024taz} that the Casimir effect is dominant not only for $ma\gg1$, but for all values of $ma$.
Hence in what follows we focus on the ground state energy arising from the Casimir effect only, which in the limit $ma\ll1$ is
    \be
    E_\text{Cas}\simeq -\frac{1}{a}\sum_{n=1}^d\frac{\Gamma(n+1)\zeta(n+1)}{2^{n-1}n\pi^{n/2}\Gamma(n/2)}~,
    \ee
as derived in Appendix \ref{sec:TDCas}. The instanton entropy keeps its generic expression (\ref{quantumS}), where the instanton action is given by eq.(\ref{Id+1}). 
The NEC and the entropy for a $d$-dimensional torus as a confining spatial geometry are then given by
    \bea \label{eq:NEC-dtunnel}
    (\rho+p)_\text{Cas}&\simeq&-\left(\frac{d+1}{d}\right)\frac{1}{a^{d+1}}\sum_{n=1}^d\frac{\Gamma(n+1)\zeta(n+1)}{2^{n-1}n\pi^{n/2}\Gamma(n/2)}~,\\
    S_\text{tun}&=&\ln(m\beta)-\frac{2m^3a^d}{\lambda}+\frac{1}{2}\ln\left(\frac{12m^3a^d}{\pi\lambda}\right)~,
    \eea
leading to the rate of change in the entropy
\be
r_{tun}\equiv a\frac{\partial S_{tun}}{\partial a}=-\frac{2d}{\lambda}m^3a^d+\frac{d}{2}~.
\ee

\subsubsection{Finite size effects in a higher dimensional sphere}

An alternative theory exhibiting tunneling was considered in \cite{Alexandre:2024gat}, 
with a massless field on a $d$-dimensional sphere $\mathbb{S}^d$ and a negative coupling to the curvature. 
In this situation, it is shown in \cite{Alexandre:2024gat} that tunnelling can actually dominate over the Casimir effect, depending on the value of the non-minimal coupling. 

The spatial volume and the curvature are
    \be 
    V=\frac{2\pi^{(d+1)/2}a^d}{\Gamma\big((d+1)/2\big)}\quad,\quad R=\frac{d(d-1)}{a^2}~,
    \ee
where $a$ is the radius of the $d$-dimensional sphere. We consider then the following action
    \be
    I=V\int d\tau\left(\frac{1}{2}(\dot\phi)^2+\frac{\lambda}{24}\left(\phi^2-\frac{6|\xi|R}{\lambda}\right)^2\right)~,
    \ee
which provides an effective set of two degenerate vacua 
    \be
    v=\pm\sqrt{\frac{6|\xi|}{\lambda}R}~,
    \ee 
and the instanton action is 
    \be 
    I_\text{inst}=
    \frac{4\pi^{(d+1)/2}(2d(d-1)|\xi|)^{3/2}}{\lambda\Gamma\big((d+1)/2\big)}a^{d-3}~.
    \ee
In the present context of a fixed geometry, the only role of the term proportional to $R^2$ in the action $I$ is to lead to the finite instanton action $I_\text{inst}$.
Of course this term would be interpreted as a modified-gravity contribution in the situation of a dynamical sphere, but this is left for future work.

For simplicity, we consider here the limit $\nu a\ll 1$ (or $|\xi|\ll1$) which is similar to the one we had for the torus, and the general case can be found in \cite{Alexandre:2024gat}.
The corresponding Casimir energies in two and three dimensions are 
    \bea
    E_\text{Cas+tun}^{\mathbb{S}^2}\simeq E_\text{Cas}^{\mathbb{S}^2}&\simeq&-\frac{0.13}{a}~,\\
    E_\text{Cas+tun}^{\mathbb{S}^3}\simeq E_\text{Cas}^{\mathbb{S}^3}&\simeq&-\frac{0.19}{a}~.
    \eea
On $\mathbb{S}^2$, the NEC and the entropy are given by
    \bea \label{eq:NEC-2sphere}
    (\rho+p)_\text{Cas}^{\mathbb{S}^2}&\simeq&-\frac{3\times0.13}{8\pi a^3}~,\\
    S_\text{tun}^{\mathbb{S}^2}&=&\ln\left(\sqrt{2|\xi|R}~\beta\right)-\frac{64\pi\xi^{3/2}}{a\lambda}+\frac{1}{2}\ln\left(\frac{384\xi^{3/2}}{a\lambda}\right)~, \label{eq:entropy-2sphere}
    \eea
with dimensionless parameters $\xi$ and $a\lambda$. 
On $\mathbb{S}^3$ we have
    \bea \label{eq:NEC-3sphere}
    (\rho+p)_\text{Cas}^{\mathbb{S}^3}&\simeq&-\frac{2\times0.19}{3\pi a^4}~,\\
    S_\text{tun}^{\mathbb{S}^3}&=&\ln\left(\sqrt{2|\xi|R}~\beta\right)-\frac{96\pi^2\sqrt{3}\xi^{3/2}}{\lambda}+\frac{1}{2}\ln\left(\frac{576\pi\sqrt{3}\xi^{3/2}}{\lambda}\right)~,
    \eea
with dimensionless parameters $\xi$ and $\lambda$.
Note that, unlike the situation of the torus, the temperature-dependent term $\ln\left(\sqrt{2|\xi|R}~\beta\right)$ in the entropy depends on $a$.
The divergence at zero temperature $\beta\to\infty$ cancels out when expressing the rate of change of the entropy though: 
\bea
r_{tun}^{\mathbb{S}^2}&\equiv&a\frac{\partial S_\text{tun}^{\mathbb{S}^2}}{\partial a}
=-\frac{3}{2}+\frac{64\pi \xi^{3/2}}{a\lambda}\\
r_{tun}^{\mathbb{S}^3}&\equiv&a\frac{\partial S_\text{tun}^{\mathbb{S}^3}}{\partial a}=-1~.
\eea

\subsection{Black holes in higher dimensions}

Here, we will review results for black holes in $2+1$ and $3+1$ gravity, but we also derive expressions for NEC violation. 
Black hole solutions are in general different depending on the number of dimensions, the approach, and the asymptotic behavior of spacetime. 
We don't have general dimensional results that fully incorporate backreaction.

\subsubsection{The BTZ black hole}

The lack of propagating degrees of freedom in $2+1$ gravity makes it much simpler than the $3+1$ dimensional one. However, black holes still exist in such a spacetime with a significant difference: they need to live in a spacetime with a negative cosmological constant $(\Lambda < 0)$. That solution was discovered by Bañados, Teitelboim and Zanelli (BTZ) in 1992 \cite{Banados:1992wn}. In contrast, no black hole solutions are known in flat $(\Lambda = 0)$ or de Sitter $(\Lambda > 0)$ spacetimes. Thus this analysis will be necessarily different than in other spacetime dimensions. The metric of a static classical BTZ black hole solution with mass $M$ is
\be
\label{eqn:metric}
ds^2=-f(r)^2 dt^2+f(r)^{-2} dr^2+r^2 d\phi^2 \,,
\ee
where 
\be
f(r)=\left(-M+\frac{r^2}{\ell^2}\right)^{1/2} \,.
\ee
This metric satisfies the vacuum field equations of (2+1)-dimensional general relativity with cosmological constant $\Lambda=-1/\ell^2$
\be
G_{\mu \nu}=\frac{1}{\ell^2}g_{\mu \nu} \,,
\ee
where $G_{\mu\nu}$ is the Einstein tensor. In a purely classical BTZ black hole spacetime, the NEC is by definition trivially obeyed as it is a vacuum solution of the Einstein equations. The situation is different for quantum BTZ (quBTZ) black holes.

The complete backreaction problem for quBTZ black holes has been tackled in some variants through a holographic reformulation \cite{Emparan:1999wa, Emparan:1999fd}. The holographic method of solving the semiclassical Einstein equation for a quantum-corrected black hole is through a classical bulk dual with a black hole localized on a braneworld \cite{Emparan:2002px}. 

The $C$-metric in $AdS_4$ (the bulk), is a central element of the construction in \cite{Emparan:2020znc}, which we roughly follow. It is part of the Pleba\'nski-Demia\'nski family of type $D$ metrics \cite{Plebanski:1976gy}. The question of regularity of the metric is an important one. In order to have a finite black hole in the bulk we must restrict $x$ (the extra coordinate in the bulk) to the range $0 \leq x \leq x_1$. The $x_1$ is the smallest of the positive roots of $G(x)$, the inverse bulk metric function of $x$. The $AdS_3$ radius on the brane is $\ell_3$ while the $AdS_4$ radius is 
\be
\ell_4=\left(\frac{1}{\ell^2}+\frac{1}{\ell_3^2} \right)^{-1/2} \,.
\ee
The parameter $\ell$ is directly related to the brane tension and the strength of backreaction in the dual theory.

The metric induced on the brane at $x= 0$ is
\be
ds^2=-\left(\frac{\bar{r}^2}{\ell_3^2}-8\mathcal{G}_3 M-\frac{\ell F(M)}{\bar{r}} \right) d\bar{t}^2+\left(\frac{\bar{r}^2}{\ell_3^2}-8\mathcal{G}_3M-\frac{\ell F(M)}{\bar{r}}\right)^{-1} d\bar{r}^2+\bar{r}^2d\bar{\phi}^2 \,,
\ee
where the coordinates are rescaled as 
\be
t=\Delta \bar{t}\,, \quad \phi=\Delta \bar{\phi}\,, \quad r=\frac{\bar{r}}{\Delta} \,,
\ee
and $\Delta=2x_1/(3+x_1^2)$. The rescaled Newton's constant is $\mathcal{G}_3 =(\ell_4/\ell) G_3=(1/2\ell) G_4$ up to order $\mathcal{O}(\ell/\ell_3)$. The function $F(M)$ is given by
\be
F(M)=8\frac{1+x_1^2}{(3+x_1^2)^3} \,,
\ee
so it depends on the value of $x_1$.

Following the holographic dictionary, the metric induced on the brane solves the
semiclassical Einstein equation. Therefore the holographic CFT stress tensor is the stress-tensor entering the equation. Solving the equation gives that stress tensor which, as an approximation is
\be
\label{eq:Tbtz}
\langle T^\mu_\nu \rangle=\frac{c}{8\pi} \frac{F(M)}{\bar{r}^3} \text{diag} \{1,1,-2\}+\mathcal{O}\left(\left(\frac{cG_3}{\ell_3}\right)^2\right) \,,
\ee
where $c$ is the central charge of the CFT. There are different branches of black hole solutions depending on the value of $x_1$ and so $F(M)$ \cite{Emparan:2020znc}. 

The NEC has been computed in \cite{Kolanowski:2023hvh}. Let $\gamma$ be an affinely-parametrized null geodesic and $u =u^t \partial t +u^r \partial_r +u^\phi \partial_\phi$ a null vector tangent to $\gamma$. Then 
\be
\label{eqn:NECBTZ}
\langle T_{\mu \nu} \rangle u^\mu u^\nu =-\frac{c}{8\pi} \frac{F(M)}{\bar{r}}(u^\phi)^2 \,.
\ee
It is clear that the NEC is violated in this case. To find its dependence with the temperature, we need the exact form of the parameters on the horizon. The horizon is on $r_+$ so $\bar{r}_+=\Delta r_+$. Following \cite{Emparan:2020znc} it is useful to introduce the following variables
\be
z=\frac{\ell_3}{r_+ x_1}\geq 0\,, \qquad \nu=\frac{\ell}{\ell_3} \,.
\ee
Then we have
\be
\bar{r}_+=\Delta r_+=2\ell_3 \frac{z(1+\nu z)}{1+3z^2-\nu z^3} \,.
\ee

\begin{figure}
    \centering
    \includegraphics[width=0.75\linewidth]{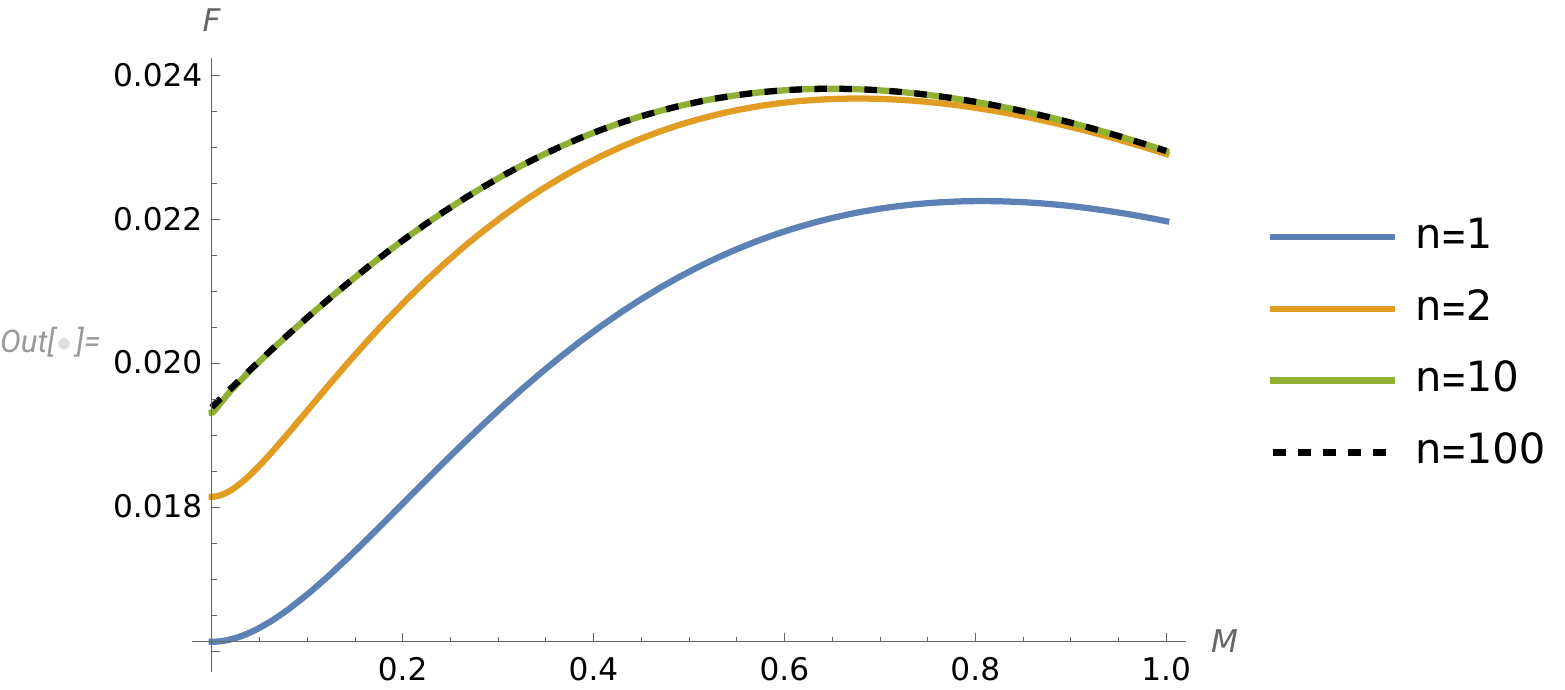}
    \caption{$F(M)$ as given by \eqref{eq:FM} for different values of $n$. As we can see, the series converges very quickly.}
    \label{fig:FM}
\end{figure}

The renormalised stress-energy for a free conformal scalar field in the BTZ black hole has the same form as in \eqref{eq:Tbtz} \cite{Emparan:2020znc}, but with $F(M)$ given by
\be
\label{eq:FM}
F(M)=\frac{(8G_3M)^{3/2}}{2\sqrt{2}}\sum_{n=1}^{\infty}\frac{\cosh(2 n\pi\sqrt{8G_3M})+3}{(\cosh(2 n\pi \sqrt{8 G_3M})-1)^{3/2}} ~.
\ee
Fig.~\ref{fig:FM} shows that a few terms of the sum are enough to give an accurate value of $F(M)$.
For $G_3M\ll 1$, and introducing a cutoff $\bar{N}$ for the total number of terms in the sum for the $M^2$ term of the Taylor expansion, we have
\be
F(M)\simeq \frac{\zeta(3)}{2\pi^3}+\frac{7\pi M^2}{240}\frac{\bar{N}(\bar{N}+1)}{2}+\mathcal{O}\left(M^4\right) \,,
\ee
where $\zeta$ is the Riemann zeta function. We write the mass $M$ in terms of $z$ and $\nu$
\be
M=\frac{\sqrt{1+\nu^2}}{G_3}\frac{z^2(1-\nu z^3)(1+\nu z)}{(1+3z^2+2\nu z^3)^2}\,.
\ee
If we want the first order quantities we can take $\nu \to 0$ meaning that $\ell \ll \ell_3$, so the backreaction effects are minimal. For $G_3M\ll 1$, the NEC in \eqref{eqn:NECBTZ} is
\bea
\label{eq:NECBTZfinal}
NEC &\simeq&-\frac{c}{16\ell_3^2\pi^2}\frac{1}{T}\left(\frac{\zeta(3)}{2\pi^3}+\frac{7\pi^5 \bar{N}(\bar{N}+1)}{1920}\frac{\ell_3^4}{G_3^2}T^4\right)\nonumber\\
&\simeq & -\frac{7\pi^3 \bar{N}(\bar{N}+1)}{2^{10}30}\frac{c\ell_3^2}{G_3^2}T^3\,.
\eea
In the last step we use the normalization of the central charge in \cite{Emparan:2020znc}, $c=\frac{\ell_4^2}{G_4}=\frac{\ell_4}{2G_3}$ and since $\ell_3>\ell_4$, we have that $\frac{\ell_3^2}{G_3^2}\ge 4 c^2$. The last approximation is valid in the limit $G_3M\ll 1$ we are considering. 

We can also compute the entropy in the regime of small $\nu$ \cite{Panella:2024sor}
\be \label{eq:entropy-BTZ-smallnu}
S=\frac{\pi \ell_3 z}{G_3(1+3z^2)}=\frac{\pi^2 \ell_3^2}{G_3} T \,.
\ee
So the entropy is proportional to the temperature in this case.

\subsubsection{$3+1$-dimensions}

Moving to $3+1$-dimensions, there are no computations that take into account fully the effects of backreaction. 
There are computations that calculate the quantum stress tensor on a black hole background either semi-analytically \cite{Visser:1996iw, Visser:1996iv, Visser:1997sd} or numerically \cite{Levi:2016quh}. 

Starting from the work of Visser, he computed semi-analytically the stress-energy tensor of a conformally coupled scalar in a Schwarzschild background. We first focus on the Unruh vacuum \cite{Visser:1997sd} in order to compare the result with the numerical approach.

The stress-tensor is assumed to be spherically symmetric. Then we know that for any $s$-wave quantum state $\psi$
\be
\langle \psi | T^{\mu \nu} | \psi \rangle \equiv \left[ \begin{array}{cccc} 
\rho & f & 0 & 0 \\
f & -\tau & 0 & 0 \\
0 & 0 & p & 0 \\
0 & 0 & 0 & p \end{array}\right]  \,.
\ee
The quantities  $\rho$, $\tau$, $f$, and $p$ are at the moment just functions of $r$, $G$ and $M$. 
Using the decomposition of Christensen and Fulling \cite{Christensen:1977jc}, Visser decomposes the stress-energy tensor in four conserved quantities. 
In addition, the trace of stress-tensor is known for conformally coupled fields (it is the conformal anomaly). 
Evaluating each conserved component individually, the functions are computed near the horizon (we are keeping the first term in the expansion)
\be
\rho(y)=-f_0 \frac{y^2}{1-y} \,, \quad \tau(y)=f_0 \frac{y^2}{1-y} \,, \quad f(y)=f_0 \frac{y^2}{1-y} ~,~\mbox{where}~~ y\equiv \frac{2GM}{r}~.
\ee
The constant $f_0$ is expressing the outgoing flux of energy. As at asymptotic spatial infinity the stress-energy in Unruh vacuum should  look like that of an outgoing flux of positive radiation, we have
\be
f_0=\frac{3}{160 \pi^2} \frac{1}{M^4} ~.
\ee
Looking at the null energy for the outward pointing radial null direction we have
\be
NEC_U=\rho-\tau-2f=-\frac{3}{40\pi^2} \frac{1}{M^4} \frac{y^2}{1-y} \,.
\ee
While this result seems to imply that the NEC is violated everywhere, we should recall that we used the near-horizon approximation. 
However, using a numerical approach Visser \cite{Visser:1997sd} showed that indeed in the Unruh vacuum the NEC is violated everywhere outside of the horizon. 
Using the known relation between mass and Hawking temperature, we have
\be
T=\frac{1}{8\pi G M} ~,
\ee
where the Boltzmann constant is set to 1, and 
\be 
NEC_U = -\frac{3}{40\pi^2} \frac{y^2}{1-y} T^4 \,.
\ee
It is clear that this expression diverges on the horizon ($z=1$) but it is regular everywhere else. For example for $r=3GM$ we have
\be \label{eq:NEC-U-1+3}
NEC_U \simeq -0.01 T^4 \,.
\ee
A similar result is derived by purely numerical methods by Levi and Ori \cite{Levi:2016quh}. 
They consider minimally coupled scalar fields instead of conformally coupled. They find NEC violation for $2GM < r \lesssim 4.9 GM$. 
For a different null geodesic, the circular trajectory at $r=3GM$ they find\footnote{See also \cite{Kontou:2023ntd} for a similar analysis.}
\be
NEC_U \simeq -0.1 T^4 \,.
\ee
Despite the difference in constants the dependence on the black hole temperature is the same.

For the Hartle Hawking vacuum \cite{Visser:1996iw}, the stress-energy tensor is
\be
\langle HH | T^{\mu\nu} | HH \rangle \equiv \left[ \begin{array}{cccc} 
-\rho & 0 & 0 & 0 \\
 0 & -\tau & 0 & 0 \\
0 & 0 & p & 0 \\
0 & 0 & 0 & p \end{array}\right]  \,.
\ee 
The NEC reduces to the constraints $\rho(r)-\tau(r)\ge 0$ and $\rho(r)+p(r)\ge 0$. Using Page's approximation, $\rho-\tau$ is always positive outside the horizon. On the other hand, $\rho(r)+p(r)$ is negative from $r=2GM$ to $r\sim 2.18994 GM$ \cite{Visser:1996iw}. At the horizon, 
\be
\label{eq:NEC-HH-1+3}
NEC_H=\rho(r)+p(r)=-24 p_{\infty}=\frac{-4}{15(16\pi)^2(2M)^4}\sim -T^4 \,.
\ee
Similar results are derived for the Boulware vacuum. 
Thus, we can confidently assume that computing the stress-energy tensor of scalar field on a Schwarzschild background gives that result.

For the black hole entropy we can consider that the first term in a spherically symmetric black hole would be the classical term. 
Then it is proportional to the area of the horizon $A$, the famous Bekenstein–Hawking result
\be
S= \frac{A}{4G} ~,
\ee
which, in terms of the temperature, reads
\be \label{eq:Entropy-Schw}
S=\frac{1}{16 \pi GT^2} ~.
\ee
However, this result does not incorporate backreaction and it is highly dependent on the symmetry and asymptotic behavior of the spacetime. 
For example Kerr black holes while they have an entropy proportional to the area of the horizon, they have a different temperature as in general the temperature of the black hole is proportional to its surface gravity. 

Another example is the $3+1$ Schwarzschild/AdS black hole \cite{Hawking:1982dh,Socolovsky:2017nff}. 
The entropy is still proportional to the area $S=\pi r_h^2/G$ where $r_h$ is the horizon radius. 
The mass is related to $r_h$ by $M=\frac{r_h}{2 G}\left(1+\frac{r_h^2}{\ell_4^2}\right)$, and the Hawking temperature is related to $r_h$ according to
\be
T=\frac{1}{4\pi r_h}\left(1+\frac{3 r_h^2}{\ell_4^2}\right)\,.
\ee
For $r_h\gg \ell_4$, i.e. large black holes, 
\be T\simeq \frac{3 r_h}{4\pi \ell_4^2}\,, \qquad r_h\simeq\frac{4\pi \ell_4^2 T}{3}\, \qquad  \text{and} \qquad S\simeq \frac{16\pi^3 \ell_4^4}{9G}T^2. 
\ee
For the small black hole limit, $r_h\ll \ell_4$
\be 
 T\simeq \frac{1}{4\pi r_h}\,, \qquad  r_h\simeq \frac{1}{4\pi T}\, , 
\ee and consequently\footnote{
Given the relationship between $M$ and $r_h$, we have $dM/d r_h>0$. The specific heat is $C=\frac{dM}{d T}=\frac{dM}{dr_h}\frac{dr_h}{dT}$. 
For large black holes $dr_h/dT>0$ while for small ones $dr_h/dT<0$, i.e. $C<0$ and the black  hole is unstable.}
\be \label{eq:AdS-S-largeb}
S\simeq\frac{1}{16 \pi G T^2}~.
\ee

\subsection{Comparison in higher dimensions}

We have explored possible extensions of the analogy between finite-volume QFT and black hole thermodynamics to higher dimensions. In 2+1 dimensions, we found a compelling example that preserves the mapping $T\leftrightarrow a^{-1}$: a scalar field undergoing tunnelling on a 2-sphere and a semiclassical BTZ black hole. Due to curvature effects, the tunnelling entropy for the scalar field presents a very special scaling with the length $S\propto 1/a $ in the limit $| \xi| \ll 1$ [see eq. \eqref{eq:entropy-2sphere}].  The semiclassical corrections for the BTZ black hole have been computed using holographic methods. In the limit $\nu \ll 1$, the entropy scales as $S\propto T$ [see eq. \eqref{eq:entropy-BTZ-smallnu}], making the analogy manifest. The NEC is violated in both cases. For for $m a\ll 1$, we find $\rho+p \propto -a^{-3}$ in the finite QFT setup \eqref{eq:NEC-2sphere}, while for $\nu \ll 1$ and $G_3M\ll 1$ the null energy condition scales as $NEC \propto -T^3$ for the semiclassical BTZ black hole \eqref{eq:NECBTZfinal}. We note that based on the analogy $T\leftrightarrow a^{-1}$ the limits $m a\ll 1$ and $G_3M\ll 1$ are similar. \\

The situation in $3+1$-dimensions, is more subtle. For the black hole side, we considered scalar quantum fluctuations in Schwarzschild and Schwarzschild anti de Sitter (S-AdS) black holes, while for the finite volume QFT side, we considered a scalar field undergoing tunnelling on a 3-sphere and on a 3-torus. Although we could match the null energy condition for the quantum fluctuations in Schwarzschild [evaluated at $r=3GM$ for the Unruh vacuum, eq.~\eqref{eq:NEC-U-1+3} and on the horizon for the Hartle-Hawking vacuum, eq.~\eqref{eq:NEC-HH-1+3}] and in both the 3-torus and the 3-sphere [assuming $ma \ll 1$; see eqs. \eqref{eq:NEC-dtunnel} and \eqref{eq:NEC-3sphere}], we did not find a finite-volume QFT analogue that matches the black hole thermodynamic behaviours in terms of entropy scaling. We note that the entropy of 3+1-dimensional black holes varies significantly with spacetime asymptotics, and backreaction effects are harder to control [see \eqref{eq:Entropy-Schw} for Schwarzschild and \eqref{eq:AdS-S-largeb} for S-AdS in the $\ell_4\gg GM$ regime].
Our analysis in $3+1$-dimensions suggests that more work has to be done to find the correct analogue systems.

\begin{table}
	\centering
	\begin{tabular}{ccc}
		\hline\\[-10pt]
		\quad System \quad &
		\quad NEC violation \quad &
		\quad Entropy rate \quad \\[5pt]
		\hline
		\hline \\[-5pt]
        \quad
        Tunneling in $2+1$ flat spacetime &
        $ -\dfrac{\pi^2+3\zeta(3)}{4\pi \textcolor{red}{a^3}}$ &
        $1-\dfrac{4m^3 a^2}{\lambda}$\\ [20pt]
        Tunneling in $2+1$ curved spacetime &
        $ -\dfrac{3\times 0.13}{8\pi \textcolor{red}{a^3}}$ &
        $-\dfrac{3}{2}+\dfrac{64\pi \xi^{3/2}}{\textcolor{red}{a}\lambda}$\\ [20pt]
       \quad quBTZ black hole & $-\dfrac{7\pi^3 \bar{N}(\bar{N}+1)}{2^{10}30} \dfrac{c\ell_3^2}{G_3^2} \textcolor{red}{T^3}$ &  $\dfrac{\pi \ell_3^2}{G_3}\textcolor{red}{T}$ \\ [15pt] \hline \\ [-5pt]
       \quad
       Tunneling in $3+1$ flat spacetime &
        $ -\dfrac{30\pi^2+2\pi^3+90\zeta(3)}{135\pi \textcolor{red}{a^4}}$ &
        $\dfrac{3}{2}-\dfrac{6m^3 a^3}{\lambda}$\\ [20pt]
        Tunneling in $3+1$ curved spacetime &
        $ -\dfrac{2\times 0.19}{3\pi \textcolor{red}{a^4}}$ &
        $-1$\\ [20pt]
       \quad Schwarzschild black hole & $-\#\textcolor{red}{T^4}$ & ? \\ [15pt]
       \hline
\end{tabular}
\caption{Summary of some preliminary results of the comparison in higher dimensions. The red highlights the comparison.}
\end{table}

\section{Conclusions}

Motivated by the description of Hawking radiation in terms of NEC violation, 
we mapped a few aspects of black hole thermodynamics with finite volume effects in QFT arising from tunnelling in a confined space. 
Both descriptions are mainly done in $1+1$-dimensions, however we also presented some evidence in higher dimensions for the same analogy. 
They feature similar behaviours in terms of energy and entropy, 
when one identifies the inverse of the BH temperature with the finite size of the confining space for a scalar field. 
The origin of the mapping is the presence of boundaries in both systems, either in the form of a horizon for the BH, 
or in the form of periodic boundary conditions in QFT.

We focused the comparison on the regime $ma\propto M/T\lesssim1$, and not on the regime $ma\propto M/T\gg1$.
In the latter case, the non-trivial effects vanish exponentially with $a$ for tunnelling, whereas for the black hole they vanish as a power law with $T$. 
In this example, the mapping we discussed does not hold and 
a more thorough discussion is necessary to include all the regimes in this study, which is left for a future work.

One possibility is the approach described in \cite{Alexandre:2024gat}, where
a massless scalar field on a $d$-sphere is considered, and a non-minimal coupling to curvature is introduced,
which provides an effective mass. The latter depends on the curvature, such that the action of the instanton is 
not proportional to the volume, and therefore tunnelling is not suppressed exponentially with the volume.
In particular, for $d=3$ NEC violation varies as $1/a$ for all values of $a$, and such a power law dependence is more likely to match the black hole description. 
This approach does not modify the picture presented here, but further studies involving $d\ge2$ space dimensions are necessary.

An essential point in our study is the presence of an environment which is necessary to justify the static regimes we study. For tunnelling, this environment fixes the spatial period, whereas for the black hole it plays the role of a heat source and fixes the temperature. Removing this environment could also be an avenue to explore, in which case the equilibrium assumption is no longer valid; since the black hole evaporates and the confining space for the scalar field is modified by energetics of NEC violation. 

Another interesting connection point that was not discussed in this work, is that Hawking radiation can be studied as tunneling of particles through the black hole horizon \cite{Parikh2000}. Then both systems can be viewed as tunneling, one on a flat background with a finite volume and one on a curved background with an infinite volume. This connection could be explored further in future work, especially in the $3+1$ dimensional case. 

Given the above analogies, we hope to find a more formal description for this mapping in light of the AdS/CFT correspondence,
although both systems here have the same dimensionality. 
Hopefully such a mapping could be fully extended to 3+1 dimensions. As discussed in Sec.~\ref{sec:higherd}, this presents unique challenges as the backreaction has not been fully computed and the entropy dependence on the temperature varies with the kind of black holes. Establishing a concrete analogy for $3+1$ dimensional systems would be an important advancement in the field and might be relevant to either astrophysics or analogue condensed matter systems.

\section*{Acknowledgments}
The authors would like to thank Dionysios Anninos, Jose Navarro-Salas, and Andrew Svesko for useful discussions. 

% TODO: include author contributions
% \paragraph{Author contributions}
% This is optional. If desired, contributions should be succinctly described in a single short paragraph, using author initials.

% TODO: include funding information
\paragraph{Funding information}
The Science and Technology Facilities Council supports the work of JA and DB (grant No.  STFC-ST/X000753/1), the Engineering and Physical Sciences Research Council supports the work of JA (grant No. EP/V002821/1) and the Leverhulme Trust supports the work of JA and SP (grant No. RPG-2021-299). DPS is supported by the EPSRC studentship grant  EP/W524475/1. For the purpose of Open Access, the authors have applied a CC BY public copyright licence to any Author Accepted Manuscript version arising from this submission.

\begin{appendix}
\numberwithin{equation}{section}

\section{Casimir effect for static saddle points}\label{Casimir}

This appendix is based on the book \cite{Bordag}, and we consider periodic boundary conditions in space. 
We first show the derivation for vanishing temperature (limit $\beta\to\infty$),
for which the ground state energy contains an ultraviolet divergence. We then show how to include finite-temperature effects, which do not
introduce new divergences.

Evaluating the individual connected graph generating functional $W[\phi_i]$ \eqref{eqn:W} at the static saddle points $\phi_s=\pm1$ for vanishing source $j=0$ yields
\be \label{Wstatappendix}
W[\phi_s]\equiv
W_\text{stat}(a,\beta)=
\frac{1}{2}\sum_{n\in\mathbb{Z}}\sum_{l\in\mathbb{Z}}\ln\left(\frac{\nu_l^2+\omega_n^2}{\nu_l^2}\right)~,
\ee
where $\omega_n=\sqrt{m^2+k_n^2}$, and
\be 
\nu_l=\frac{2\pi l}{\beta}~~,~~
k_n=\frac{2\pi n}{a}~.
\ee
The origin of energies is chosen in such a way as to recover the usual sum of ground state energies of harmonic oscillators, 
at zero temperature.

\subsection{Zero temperature}

In the limit of zero temperature the summation over Matsubara modes becomes an integral
\be\label{Wstat1d}
\lim_{\beta\to\infty}\big\{W[\phi_s]\big\}=
\frac{\beta}{2}\sum _{n\in\mathbb{Z}} \int_{-\infty}^\infty \frac{d \nu}{2 \pi} \ln\left(\frac{\nu^2+\omega_n^2}{\nu^2}\right)
=\frac{\beta}{2}\sum _{n\in\mathbb{Z}}\omega_n~.
\ee
Using the Abel-Plana formula
\be \label{Abel-Plana}
\sum_{n\in\mathbb{N}} F(n)\equiv
-\frac{1}{2}F(0)
+\int_0^\infty F(t)dt
+i\int_0^\infty\frac{dt}{e^{2\pi t}-1}\Big[F(it)-F(-it)\Big]~,
\ee
the sum over the frequencies can be expressed as
\be \label{Frequencysum}
\sum_{n\in\mathbb{Z}} \omega(n)=
2a\Lambda^2
+2i\int_0^\infty\frac{dt}{e^{2\pi t}-1}\Big[\omega(it)-\omega(-it)\Big]~,
\ee
where the ultraviolet divergence is
\be
\Lambda^2\equiv
\frac{m^2}{2\pi}\int_0^\infty \sqrt{t^2+1}~dt~.
\ee 
By considering the principle branch $z\in]-\infty,0]$ of $\ln(z)$, we have
\be 
\omega(it)-\omega(-it)=\frac{4\pi i}{a}\sqrt{t^2-\mu^2}\theta(t-\mu)~,
\ee 
where $\mu\equiv ma/2\pi$. The individual connected graph generating functional for static saddle points therefore reads,
in the limit of zero temperature,
\bea
\lim_{\beta\to\infty}\big\{W[\phi_s]\big\}&=&
a\beta \Lambda^2
-\beta\frac{4\pi}{a}\int_\mu^\infty\frac{dt}{e^{2\pi t}-1}\sqrt{t^2-\mu^2}\\
&=& \beta a\Lambda^2
-\beta\frac{m^2a}{\pi}\int_1^\infty\frac{du}{e^{mau}-1}\sqrt{u^2-1}~.
\eea

\subsection{Finite-temperature corrections}

In order to calculate thermal corrections to the individual connected graph generating functional for static saddle points, 
we employ the zeta-function regularisation and write the expression \eqref{Wstat} as
\be 
W_\text{stat}(a,\beta,s)=
-\frac{1}{2}\frac{\partial }{\partial s}\left(\sum _{l\in\mathbb{Z}} \sum _{n\in\mathbb{Z}}
(\beta a)^{-s}\left(\nu_l^2+\omega_n^2\right)^{-s}\right)~,
\ee
where the ultraviolet divergence is turned into a divergence in the limit $s\to0$. The above can be written in terms of a parametric integral
\be \label{Fint}
W_\text{stat}(a,\beta,s)
=-\frac{1}{2}\frac{\partial }{\partial s}\left(\int _0^{\infty }\frac{dt}{t}\frac{t^s}{\Gamma(s)}
\sum _{l\in\mathbb{Z}} \sum _{n\in\mathbb{Z}}
\e^{-t\beta a\left(\nu_l^2+\omega_n^2\right)}\right)~.
\ee
From the Poisson summation formula, one can derive the following identity
\be 
\sum_{l\in\mathbb{Z}} \e^{-z l^2}=\sqrt{\frac{\pi}{z}} \sum_{l\in\mathbb{Z}} \e^{-\pi^2 l^2 / z}~,
\ee
which, when applied to the Matsubara sum in eq. \eqref{Fint} with $z=\beta a t(2\pi T)^2$, leads to
\be \label{FreeEnergy2}
W_\text{stat}(a,\beta,s)
=-\frac{\beta}{2}\frac{\partial }{\partial s}\left(\sum _{l\in\mathbb{Z}}\int _0^{\infty }\frac{dt}{t}\frac{t^s}{\Gamma(s)\sqrt{4\pi\beta a t}}
\sum _{n\in\mathbb{Z}}\e^{-\frac{l^2\beta^2}{4  \beta a t}-  \beta a t\omega_n^2}\right)~.
\ee 
The ultraviolet divergence is contained within the temperature-independent integral for $l=0$. We thus make the following decomposition of eq. \eqref{FreeEnergy2}
\be \label{FreeEnergyT2}
W_\text{stat}(a,\beta,s)=
\lim_{\beta\to\infty}\big\{W_\text{stat}(a,\beta,s)\big\}
+W_\text{stat}^T(a,\beta)~,
\ee
where the temperature independent part is calculated in the previous section, and the temperature-dependent part is given by
\be\label{FreeEnergyT}
W_\text{stat}^T(a,\beta)\equiv
-\frac{\beta}{\sqrt{4\pi  \beta a}}\sum _{l\in\mathbb{N}}\int _0^{\infty }\frac{dt}{t^{3/2}}\sum _{n\in\mathbb{Z}}
\e^{-\frac{l^2\beta^2}{4  \beta a t}-  \beta a t\omega_n^2}~.
\ee
Note that the regulator has been removed from eq. \eqref{FreeEnergyT} using
\be
\underset{s\to 0}{\lim}\left[\frac{\partial }{\partial s}\frac{f(s)}{\Gamma (s)}\right]=f(0)~.
\ee
The integral and summation over $l$ in eq. \eqref{FreeEnergyT} can be then evaluated, leading to
\be \label{FreeEnergyDis}
W_\text{stat}^T(a,\beta)=
\sum_{n\in\mathbb{Z}}\ln\left(1-\e^{-\beta\omega_n}\right)~.
\ee

\section{Casimir energy of a $d$-dimensional Torus}\label{sec:TDCas}

The results presented here were derived independently in \cite{Ai:2024taz}.\\

Consider a massive scalar field on a $d$-dimensional Torus $\mathbb{T}^d$ with corresponding boundary conditions
\be
\phi(t,x_1,\ldots,x_d)=\phi(t,x_1+a_1,\ldots,x_D+a_d)~,
\ee
and volume
    \be 
    V_{\mathbb{T}^d}=\prod_{i=1}^d a_i~.
    \ee 
The zero temperature vacuum energy density is represented as a sum over ground state energies of harmonic oscillators
\be\label{TDEVac}
E_{\mathbb{T}^d}(m)=
\frac{1}{2}\sum _{n_1}\dots\sum _{n_d}
\omega(m;n_1,\ldots,n_d)~,
\ee
where
\be 
\omega^2(m;n_1,\ldots,n_d)=m^2+\sum_{i=1}^d \left(\frac{2\pi n_i}{a_i}\right)^2
\quad,\quad
n_i\in \mathbb{Z}~.
\ee
At this stage it is easy to see that the following recursion relation holds
\be \label{TDEVac-Rec}
E_{\mathbb{T}^d}(m)\equiv
\sum_{n_d}E_{\mathbb{T}^{d-1}}\big(\Tilde{m}(n_d)\big)
\ee 
with
\be 
\Tilde{m}^2(n_d)\equiv m^2 + \left(\frac{2\pi n_d}{a_d}\right)^2~,
\ee 
where the $n_d$ term has been absorbed into the mass. The aim will be to re-apply this recursion relation after having applied the Abel-Plana summation formula \eqref{Abel-Plana} multiple times.\\

We can pull out a factor of $2\pi/a_1$ such that \eqref{TDEVac} can be written in the form
\be \label{TDEVac-2}
E_{\mathbb{T}^d}=
\frac{\pi}{a_1}
\sum _{n_1}\dots\sum _{n_d}
\sqrt{n_1^2+\mu_1(n_2,\ldots,n_d)^2}~,
\ee
where
\be \label{muRD}
\mu_i(t_1,\ldots,t_{i-1},t_{i+1},\ldots,t_d)^2\equiv
\left(\frac{ma_i}{2\pi}\right)^2
+a_i^2\sum_{j\neq i}^{d}\left(\frac{t_j}{a_j}\right)^2~,
\ee
is a function of $d-1$ arguments. We may now apply \eqref{Abel-Plana} $d-1$ times, each time converting a summation over $n_i$ to an integral over $t_i$, until reaching the expression
\be \label{TDEVac-3}
E_{\mathbb{T}^d}=
\sum _{n_d}\left(
\frac{2^{d-1}\pi}{a_d}
\int_0^\infty \dd \mathbf{t}
\sqrt{n_d^2+\mu_d(t_1,\ldots,t_{d-1})^2}
+G_d(n_d)
\right)~,
\ee
where $\mathbf{t}\equiv (t_1,\dots,t_{d-1})$ and $G(n_d)$ is some complicated expression comprising of sums over branch-cut terms of \eqref{Abel-Plana}, generated via repeated application of the formula.
Equating \eqref{TDEVac-3} with the RHS of the recursion relation \eqref{TDEVac-Rec} and dropping the summations, we find
\be
E_{\mathbb{T}^{d-1}}\big(\Tilde{m}(n_d)\big)=
\frac{2^{d-1}\pi}{a_d}
\int_0^\infty \dd \mathbf{t}
\sqrt{n_d^2+\mu_d(t_1,\ldots,t_{d-1})^2}
+G_d(n_d)~.
\ee
The first term on the RHS is exactly the vacuum energy density of $d-1$ dimensional free space with the replacement $m\to \Tilde{m}(n_d)$, which means the second term is the corresponding Casimir energy
\be 
E_{\mathbb{T}^{d-1}}^\text{Cas}\big(\Tilde{m}(n_d)\big)\equiv
G_d(n_d)~.
\ee
We may then express \eqref{TDEVac-3} as
\be \label{TDEVac-4}
E_{\mathbb{T}^d}=
\frac{2^{d-1}\pi}{a_d}
\sum _{n_d}\int_0^\infty \dd \mathbf{t}
\sqrt{n_d^2+\mu_d(t_1,\ldots,t_{d-1})^2}
+\sum_{n_d}E_{\mathbb{T}^{d-1}}^\text{Cas}\big(\Tilde{m}(n_d)\big)~.
\ee

Continuing, we apply \eqref{Abel-Plana} one last time, yielding
\bea \label{TDEVac-4}
E_{\mathbb{T}^d}&=&
\frac{2^d\pi}{a_d}
\int_0^\infty \dd t_1\ldots\dd t_{d}
\sqrt{t_d^2+\mu_d(t_1,\ldots,t_{d-1})^2}\\
&& -\frac{2^{d+1}\pi}{a_d}
\int_0^\infty \dd t_1\ldots\dd t_{d-1}
\int_{t_d=\mu_d}^\infty
\frac{\sqrt{t_d^2-\mu_d(t_1,\ldots,t_{d-1})^2}}{e^{2\pi t_d}-1}\dd t_d\\
&&+\sum_{n_d}E_{\mathbb{T}^{d-1}}^\text{Cas}\big(\Tilde{m}(n_d)\big)~.\nonumber
\eea
The first term is the vacuum energy density of unbounded space (in correspondence with the $a\to\infty$ limit of \eqref{TDEVac}) which cancels out upon evaluating the Casimir energy. The integral over $\dd t_1\ldots\dd t_{d-1}$ can be evaluated in $d-1$ hyper-spherical coordinates and leads to the following recursion relation for the Casimir energy
\be \label{TDCas}
E_{\mathbb{T}^d}^\text{Cas}=
-\frac{m^{d+1}V_{\mathbb{T}^d}}{2^{d-1}D\pi^{d/2}\Gamma(d/2)}\int_{t=1}^\infty
\frac{(t^2-1)^{d/2}}{e^{ma_d t}-1}\dd t
+\sum_{n_d}E_{\mathbb{T}^{d-1}}^\text{Cas}\big(\Tilde{m}(n_d)\big)~.\nonumber
\ee

The summation in the final term in the above expression is highly suppressed in $n_d$ such that to a good approximation ($\sim 1\%$ deviation from the true value) one may consider only the leading order term $n_d=0$. Considering also the further simplified case of a symmetric $d$-dimensional Torus, where $a_i=a_j, \forall i,j$, the approximate Casimir energy of a $d$-dimensional torus is
    \be
    E_{\mathbb{T}^d}^\text{Cas}(m,a)\simeq
    -\sum_{n=1}^d
    \frac{m^{n+1}a^n}{2^{n-1}n\pi^{n/2}\Gamma(n/2)}\int_{t=1}^\infty
    \frac{(t^2-1)^{n/2}}{e^{mat}-1}\dd t~,
    \ee
which, in the massless limit, becomes
    \be
    E_{\mathbb{T}^d}^\text{Cas}(m\to0,a)\simeq
    -\frac{1}{a}\sum_{n=1}^d
    \frac{\Gamma(n+1)\zeta(n+1)}{2^{n-1}n\pi^{n/2}\Gamma(n/2)}~.
    \ee

\end{appendix}

\newpage

\bibliographystyle{JHEP}      % ieeetrunsrtAPS-like style for physics

\bibliography{violating_NEC_via_tunneling_in_BH}

\providecommand{\href}[2]{#2}\begingroup\raggedright\begin{thebibliography}{10}

\bibitem{Hawking:1974rv}
S.W.~Hawking, \emph{{Black hole explosions}},
  \href{https://doi.org/10.1038/248030a0}{\emph{Nature} {\bfseries 248} (1974)
  30}.

\bibitem{Curiel:2014zba}
E.~Curiel, \emph{{A Primer on Energy Conditions}},
  \href{https://doi.org/10.1007/978-1-4939-3210-8_3}{\emph{Einstein Stud.}
  {\bfseries 13} (2017) 43} [\href{https://arxiv.org/abs/1405.0403}{{\ttfamily
  1405.0403}}].

\bibitem{Kontou:2020bta}
E.-A.~Kontou and K.~Sanders, \emph{{Energy conditions in general relativity and
  quantum field theory}},
  \href{https://doi.org/10.1088/1361-6382/ab8fcf}{\emph{Class. Quant. Grav.}
  {\bfseries 37} (2020) 193001}
  [\href{https://arxiv.org/abs/2003.01815}{{\ttfamily 2003.01815}}].

\bibitem{Barcelo:2000zf}
C.~Barcelo and M.~Visser, \emph{{Scalar fields, energy conditions, and
  traversable wormholes}},
  \href{https://doi.org/10.1088/0264-9381/17/18/318}{\emph{Class. Quant. Grav.}
  {\bfseries 17} (2000) 3843}
  [\href{https://arxiv.org/abs/gr-qc/0003025}{{\ttfamily gr-qc/0003025}}].

\bibitem{Fliss:2023rzi}
J.R.~Fliss, B.~Freivogel, E.-A.~Kontou and D.P.~Santos, \emph{{Non-minimal
  coupling, negative null energy, and effective field theory}},
  \href{https://arxiv.org/abs/2309.10848}{{\ttfamily 2309.10848}}.

\bibitem{Penrose:1964wq}
R.~Penrose, \emph{{Gravitational collapse and space-time singularities}},
  \href{https://doi.org/10.1103/PhysRevLett.14.57}{\emph{Phys. Rev. Lett.}
  {\bfseries 14} (1965) 57}.

\bibitem{Hawking:1971vc}
S.W.~Hawking, \emph{{Black holes in general relativity}},
  \href{https://doi.org/10.1007/BF01877517}{\emph{Commun. Math. Phys.}
  {\bfseries 25} (1972) 152}.

\bibitem{Epstein:1965zza}
H.~Epstein, V.~Glaser and A.~Jaffe, \emph{{Nonpositivity of energy density in
  Quantized field theories}},
  \href{https://doi.org/10.1007/BF02749799}{\emph{Nuovo Cim.} {\bfseries 36}
  (1965) 1016}.

\bibitem{Bordag}
M.~Bordag, G.L.~Klimchitskaya, U.~Mohideen and V.M.~Mostepanenko,
  \emph{{Advances in the Casimir Effect}} (2009),
  \href{https://doi.org/10.1093/acprof:oso/9780199238743.001.0001}{10.1093/acprof:oso/9780199238743.001.0001}.

\bibitem{Alexandre:2022qxc}
J.~Alexandre and J.~Polonyi, \emph{{Symmetry restoration, tunneling, and the
  null energy condition}},
  \href{https://doi.org/10.1103/PhysRevD.106.065008}{\emph{Phys. Rev. D}
  {\bfseries 106} (2022) 065008}
  [\href{https://arxiv.org/abs/2205.00768}{{\ttfamily 2205.00768}}].

\bibitem{Alexandre:2023iig}
J.~Alexandre and D.~Backhouse, \emph{{Null energy condition violation:
  Tunneling versus the Casimir effect}},
  \href{https://doi.org/10.1103/PhysRevD.107.085022}{\emph{Phys. Rev. D}
  {\bfseries 107} (2023) 085022}
  [\href{https://arxiv.org/abs/2301.02455}{{\ttfamily 2301.02455}}].

\bibitem{Alexandre:2023pkk}
J.~Alexandre and S.~Pla, \emph{{Cosmic bounce and phantom-like equation of
  state from tunnelling}},
  \href{https://doi.org/10.1007/JHEP05(2023)145}{\emph{JHEP} {\bfseries 05}
  (2023) 145} [\href{https://arxiv.org/abs/2301.08652}{{\ttfamily
  2301.08652}}].

\bibitem{Alexandre:2023bih}
J.~Alexandre, K.~Clough and S.~Pla, \emph{{Tunneling-induced cosmic bounce in
  the presence of anisotropies}},
  \href{https://doi.org/10.1103/PhysRevD.108.103515}{\emph{Phys. Rev. D}
  {\bfseries 108} (2023) 103515}
  [\href{https://arxiv.org/abs/2308.00765}{{\ttfamily 2308.00765}}].

\bibitem{Ai:2024taz}
W.-Y.~Ai, J.~Alexandre, M.~Carosi, B.~Garbrecht and S.~Pla, \emph{{Double-well
  instantons in finite volume}},
  \href{https://doi.org/10.1007/JHEP05(2024)099}{\emph{JHEP} {\bfseries 05}
  (2024) 099} [\href{https://arxiv.org/abs/2402.09863}{{\ttfamily
  2402.09863}}].

\bibitem{Symanzik:1969ek}
K.~Symanzik, \emph{{Renormalizable models with simple symmetry breaking. 1.
  Symmetry breaking by a source term}},
  \href{https://doi.org/10.1007/BF01645494}{\emph{Commun. Math. Phys.}
  {\bfseries 16} (1970) 48}.

\bibitem{Coleman:1974jh}
S.R.~Coleman, R.~Jackiw and H.D.~Politzer, \emph{{Spontaneous Symmetry Breaking
  in the O(N) Model for Large N*}},
  \href{https://doi.org/10.1103/PhysRevD.10.2491}{\emph{Phys. Rev. D}
  {\bfseries 10} (1974) 2491}.

\bibitem{Iliopoulos:1974ur}
J.~Iliopoulos, C.~Itzykson and A.~Martin, \emph{{Functional Methods and
  Perturbation Theory}},
  \href{https://doi.org/10.1103/RevModPhys.47.165}{\emph{Rev. Mod. Phys.}
  {\bfseries 47} (1975) 165}.

\bibitem{Haymaker:1983xk}
R.W.~Haymaker and J.~Perez-Mercader, \emph{{Convexity of the Effective
  Potential}}, \href{https://doi.org/10.1103/PhysRevD.27.1948}{\emph{Phys. Rev.
  D} {\bfseries 27} (1983) 1948}.

\bibitem{Fujimoto:1982tc}
Y.~Fujimoto, L.~O'Raifeartaigh and G.~Parravicini, \emph{{Effective Potential
  for Nonconvex Potentials}},
  \href{https://doi.org/10.1016/0550-3213(83)90305-X}{\emph{Nucl. Phys. B}
  {\bfseries 212} (1983) 268}.

\bibitem{Bender:1983nc}
C.M.~Bender and F.~Cooper, \emph{{Failure of the Naive Loop Expansion for the
  Effective Potential in $\phi^4$ Field Theory When There Is 'Broken
  Symmetry'}}, \href{https://doi.org/10.1016/0550-3213(83)90383-8}{\emph{Nucl.
  Phys. B} {\bfseries 224} (1983) 403}.

\bibitem{Hindmarsh:1985nc}
M.~Hindmarsh and D.~Johnston, \emph{{Convexity of the Effective Potential}},
  \href{https://doi.org/10.1088/0305-4470/19/1/016}{\emph{J. Phys. A}
  {\bfseries 19} (1986) 141}.

\bibitem{Alexandre:2012ht}
J.~Alexandre and A.~Tsapalis, \emph{{Maxwell Construction for Scalar Field
  Theories with Spontaneous Symmetry Breaking}},
  \href{https://doi.org/10.1103/PhysRevD.87.025028}{\emph{Phys. Rev. D}
  {\bfseries 87} (2013) 025028}
  [\href{https://arxiv.org/abs/1211.0921}{{\ttfamily 1211.0921}}].

\bibitem{Plascencia:2015pga}
A.D.~Plascencia and C.~Tamarit, \emph{{Convexity, gauge-dependence and
  tunneling rates}}, \href{https://doi.org/10.1007/JHEP10(2016)099}{\emph{JHEP}
  {\bfseries 10} (2016) 099}
  [\href{https://arxiv.org/abs/1510.07613}{{\ttfamily 1510.07613}}].

\bibitem{Millington:2019nkw}
P.~Millington and P.M.~Saffin, \emph{{Visualising quantum effective action
  calculations in zero dimensions}},
  \href{https://doi.org/10.1088/1751-8121/ab37e6}{\emph{J. Phys. A} {\bfseries
  52} (2019) 405401} [\href{https://arxiv.org/abs/1905.09674}{{\ttfamily
  1905.09674}}].

\bibitem{Hu:2021fjq}
X.-Y.~Hu, M.~Kleban and C.~Yu, \emph{{Electric field decay without pair
  production: lattice, bosonization and novel worldline instantons}},
  \href{https://doi.org/10.1007/JHEP03(2022)197}{\emph{JHEP} {\bfseries 03}
  (2022) 197} [\href{https://arxiv.org/abs/2107.04561}{{\ttfamily
  2107.04561}}].

\bibitem{Rubakov:2014jja}
V.A.~Rubakov, \emph{{The Null Energy Condition and its violation}},
  \href{https://doi.org/10.3367/UFNe.0184.201402b.0137}{\emph{Phys. Usp.}
  {\bfseries 57} (2014) 128} [\href{https://arxiv.org/abs/1401.4024}{{\ttfamily
  1401.4024}}].

\bibitem{Easson:2024fzn}
D.A.~Easson and J.E.~Lesnefsky, \emph{{Eternal Universes}},
  \href{https://arxiv.org/abs/2404.03016}{{\ttfamily 2404.03016}}.

\bibitem{Sopova:2002cs}
V.~Sopova and L.H.~Ford, \emph{{The Energy density in the Casimir effect}},
  \href{https://doi.org/10.1103/PhysRevD.66.045026}{\emph{Phys. Rev. D}
  {\bfseries 66} (2002) 045026}
  [\href{https://arxiv.org/abs/quant-ph/0204125}{{\ttfamily
  quant-ph/0204125}}].

\bibitem{Graham:2002yr}
N.~Graham and K.D.~Olum, \emph{{Negative energy densities in quantum field
  theory with a background potential}},
  \href{https://doi.org/10.1103/PhysRevD.69.109901}{\emph{Phys. Rev. D}
  {\bfseries 67} (2003) 085014}
  [\href{https://arxiv.org/abs/hep-th/0211244}{{\ttfamily hep-th/0211244}}].

\bibitem{Graham:2005cq}
N.~Graham and K.D.~Olum, \emph{{Plate with a hole obeys the averaged null
  energy condition}},
  \href{https://doi.org/10.1103/PhysRevD.72.025013}{\emph{Phys. Rev. D}
  {\bfseries 72} (2005) 025013}
  [\href{https://arxiv.org/abs/hep-th/0506136}{{\ttfamily hep-th/0506136}}].

\bibitem{Coleman:1977py}
S.R.~Coleman, \emph{{The Fate of the False Vacuum. 1. Semiclassical Theory}},
  \href{https://doi.org/10.1103/PhysRevD.16.1248}{\emph{Phys. Rev. D}
  {\bfseries 15} (1977) 2929}.

\bibitem{Callan:1977pt}
C.G.~Callan, Jr. and S.R.~Coleman, \emph{{The Fate of the False Vacuum. 2.
  First Quantum Corrections}},
  \href{https://doi.org/10.1103/PhysRevD.16.1762}{\emph{Phys. Rev. D}
  {\bfseries 16} (1977) 1762}.

\bibitem{Kleinert:2004ev}
H.~Kleinert, \emph{{Path Integrals in Quantum Mechanics, Statistics, Polymer
  Physics, and Financial Markets}} (2004),
  \href{https://doi.org/10.1142/5057}{10.1142/5057}.

\bibitem{Cho:2022ngf}
H.-T.~Cho, J.-T.~Hsiang and B.-L.~Hu, \emph{{Quantum Capacity and Vacuum
  Compressibility of Spacetime: Thermal Fields}},
  \href{https://doi.org/10.3390/universe8050291}{\emph{Universe} {\bfseries 8}
  (2022) 291} [\href{https://arxiv.org/abs/2204.08634}{{\ttfamily
  2204.08634}}].

\bibitem{Brevik:2002gh}
I.H.~Brevik, K.A.~Milton and S.D.~Odintsov, \emph{{Entropy bounds in R x S**3
  geometries}}, \href{https://doi.org/10.1006/aphy.2002.6317}{\emph{Annals
  Phys.} {\bfseries 302} (2002) 120}
  [\href{https://arxiv.org/abs/hep-th/0202048}{{\ttfamily hep-th/0202048}}].

\bibitem{1992_Callan}
C.G.~Callan, S.B.~Giddings, J.A.~Harvey and A.~Strominger, \emph{Evanescent
  black holes}, \href{https://doi.org/10.1103/PhysRevD.45.R1005}{\emph{Phys.
  Rev. D} {\bfseries 45} (1992) R1005}.

\bibitem{1994_Fiola}
T.M.~Fiola, J.~Preskill, A.~Strominger and S.P.~Trivedi, \emph{{Black hole
  thermodynamics and information loss in two-dimensions}},
  \href{https://doi.org/10.1103/PhysRevD.50.3987}{\emph{Phys. Rev. D}
  {\bfseries 50} (1994) 3987}
  [\href{https://arxiv.org/abs/hep-th/9403137}{{\ttfamily hep-th/9403137}}].

\bibitem{1992_Russo}
J.G.~Russo, L.~Susskind and L.~Thorlacius, \emph{End point of hawking
  radiation}, \href{https://doi.org/10.1103/PhysRevD.46.3444}{\emph{Phys. Rev.
  D} {\bfseries 46} (1992) 3444}.

\bibitem{1995_Bose}
S.~Bose, L.~Parker and Y.~Peleg, \emph{{Semiinfinite throat as the end state
  geometry of two-dimensional black hole evaporation}},
  \href{https://doi.org/10.1103/PhysRevD.52.3512}{\emph{Phys. Rev. D}
  {\bfseries 52} (1995) 3512}
  [\href{https://arxiv.org/abs/hep-th/9502098}{{\ttfamily hep-th/9502098}}].

\bibitem{Cruz:1995zt}
J.~Cruz and J.~Navarro-Salas, \emph{{Solvable models for radiating black holes
  and area preserving diffeomorphisms}},
  \href{https://doi.org/10.1016/0370-2693(96)00246-8}{\emph{Phys. Lett. B}
  {\bfseries 375} (1996) 47}
  [\href{https://arxiv.org/abs/hep-th/9512187}{{\ttfamily hep-th/9512187}}].

\bibitem{1984_Jackiw}
R.~Jackiw, \emph{{Lower Dimensional Gravity}},
  \href{https://doi.org/10.1016/0550-3213(85)90448-1}{\emph{Nucl. Phys. B}
  {\bfseries 252} (1985) 343}.

\bibitem{2002_Grumiller}
D.~Grumiller, W.~Kummer and D.V.~Vassilevich, \emph{{Dilaton gravity in
  two-dimensions}},
  \href{https://doi.org/10.1016/S0370-1573(02)00267-3}{\emph{Phys. Rept.}
  {\bfseries 369} (2002) 327}
  [\href{https://arxiv.org/abs/hep-th/0204253}{{\ttfamily hep-th/0204253}}].

\bibitem{fabbri-navarro}
A.~Fabbri and J.~Navarro-Salas, \emph{{Modeling black hole evaporation}},
  Imperial College Press-World Scientific, London (2005).

\bibitem{2022_Djordjevic}
S.~Djordjevi\'c, A.~Go\v{c}anin, D.~Go\v{c}anin and V.~Radovanovi\'c,
  \emph{{Page curve for an eternal Schwarzschild black hole in a dimensionally
  reduced model of dilaton gravity}},
  \href{https://doi.org/10.1103/PhysRevD.106.105015}{\emph{Phys. Rev. D}
  {\bfseries 106} (2022) 105015}
  [\href{https://arxiv.org/abs/2207.07409}{{\ttfamily 2207.07409}}].

\bibitem{1992_Giddings}
S.B.~Giddings and A.~Strominger, \emph{{Dynamics of extremal black holes}},
  \href{https://doi.org/10.1103/PhysRevD.46.627}{\emph{Phys. Rev. D} {\bfseries
  46} (1992) 627} [\href{https://arxiv.org/abs/hep-th/9202004}{{\ttfamily
  hep-th/9202004}}].

\bibitem{1992_Strominger}
A.~Strominger, \emph{{Faddeev-Popov ghosts and (1+1)-dimensional black hole
  evaporation}}, \href{https://doi.org/10.1103/PhysRevD.46.4396}{\emph{Phys.
  Rev. D} {\bfseries 46} (1992) 4396}
  [\href{https://arxiv.org/abs/hep-th/9205028}{{\ttfamily hep-th/9205028}}].

\bibitem{1992_Harvey}
J.A.~Harvey and A.~Strominger, \emph{{Quantum aspects of black holes}},  in
  \emph{{Spring School on String Theory and Quantum Gravity (To be followed by
  Workshop on String Theory 8-10 Apr)}}, pp.~175--223, 1993
  [\href{https://arxiv.org/abs/hep-th/9209055}{{\ttfamily hep-th/9209055}}].

\bibitem{Mikovic:1997mz}
A.R.~Mikovic and V.~Radovanovic, \emph{{One loop effective action for a generic
  2-D dilaton gravity theory}},
  \href{https://doi.org/10.1016/S0550-3213(97)00474-4}{\emph{Nucl. Phys. B}
  {\bfseries 504} (1997) 511}
  [\href{https://arxiv.org/abs/hep-th/9704014}{{\ttfamily hep-th/9704014}}].

\bibitem{1977_Christensen}
S.M.~Christensen and S.A.~Fulling, \emph{Trace anomalies and the hawking
  effect}, \href{https://doi.org/10.1103/PhysRevD.15.2088}{\emph{Phys. Rev. D}
  {\bfseries 15} (1977) 2088}.

\bibitem{1994_Myers}
R.C.~Myers, \emph{{Black hole entropy in two-dimensions}},
  \href{https://doi.org/10.1103/PhysRevD.50.6412}{\emph{Phys. Rev. D}
  {\bfseries 50} (1994) 6412}
  [\href{https://arxiv.org/abs/hep-th/9405162}{{\ttfamily hep-th/9405162}}].

\bibitem{2023_Hirano}
S.~Hirano, \emph{{Island formula from Wald-like entropy with backreaction}},
  \href{https://doi.org/10.1007/JHEP02(2024)125}{\emph{JHEP} {\bfseries 02}
  (2024) 125} [\href{https://arxiv.org/abs/2310.03416}{{\ttfamily
  2310.03416}}].

\bibitem{1994_Iyer}
V.~Iyer and R.M.~Wald, \emph{Some properties of the noether charge and a
  proposal for dynamical black hole entropy},
  \href{https://doi.org/10.1103/PhysRevD.50.846}{\emph{Phys. Rev. D} {\bfseries
  50} (1994) 846}.

\bibitem{2022_Yu}
M.-H.~Yu and X.-H.~Ge, \emph{{Entanglement islands in generalized
  two-dimensional dilaton black holes}},
  \href{https://doi.org/10.1103/PhysRevD.107.066020}{\emph{Phys. Rev. D}
  {\bfseries 107} (2023) 066020}
  [\href{https://arxiv.org/abs/2208.01943}{{\ttfamily 2208.01943}}].

\bibitem{2021_Pedraza}
J.F.~Pedraza, A.~Svesko, W.~Sybesma and M.R.~Visser, \emph{{Semi-classical
  thermodynamics of quantum extremal surfaces in Jackiw-Teitelboim gravity}},
  \href{https://doi.org/10.1007/JHEP12(2021)134}{\emph{JHEP} {\bfseries 12}
  (2021) 134} [\href{https://arxiv.org/abs/2107.10358}{{\ttfamily
  2107.10358}}].

\bibitem{Hayward:1994dw}
J.D.~Hayward, \emph{{Entropy in the RST model}},
  \href{https://doi.org/10.1103/PhysRevD.52.2239}{\emph{Phys. Rev. D}
  {\bfseries 52} (1995) 2239}
  [\href{https://arxiv.org/abs/gr-qc/9412065}{{\ttfamily gr-qc/9412065}}].

\bibitem{Visser:1996ix}
M.~Visser, \emph{{Gravitational vacuum polarization. 3: Energy conditions in
  the (1+1) Schwarzschild space-time}},
  \href{https://doi.org/10.1103/PhysRevD.54.5123}{\emph{Phys. Rev. D}
  {\bfseries 54} (1996) 5123}
  [\href{https://arxiv.org/abs/gr-qc/9604009}{{\ttfamily gr-qc/9604009}}].

\bibitem{Widom:1998pn}
A.~Widom, E.~Sassaroli, Y.N.~Srivastava and J.~Swain, \emph{{The Casimir effect
  and thermodynamic instability}},
  \href{https://arxiv.org/abs/quant-ph/9803013}{{\ttfamily quant-ph/9803013}}.

\bibitem{Sassaroli:1998ug}
E.~Sassaroli, Y.N.~Srivastava, J.~Swain and A.~Widom, \emph{{The Dynamical and
  static Casimir effects and the thermodynamic instability}},  in \emph{{ITAMP
  Topical Group on Casimir Forces Atomic and molecular physics.}}, 3, 1998
  [\href{https://arxiv.org/abs/hep-ph/9805479}{{\ttfamily hep-ph/9805479}}].

\bibitem{Alexandre:2024gat}
J.~Alexandre and D.~Backhouse, \emph{{Tunnelling and the Casimir effect on a
  $D$-dimensional sphere}},  \href{https://arxiv.org/abs/2408.17189}{{\ttfamily
  2408.17189}}.

\bibitem{Banados:1992wn}
M.~Banados, C.~Teitelboim and J.~Zanelli, \emph{{The Black hole in
  three-dimensional space-time}},
  \href{https://doi.org/10.1103/PhysRevLett.69.1849}{\emph{Phys. Rev. Lett.}
  {\bfseries 69} (1992) 1849}
  [\href{https://arxiv.org/abs/hep-th/9204099}{{\ttfamily hep-th/9204099}}].

\bibitem{Emparan:1999wa}
R.~Emparan, G.T.~Horowitz and R.C.~Myers, \emph{{Exact description of black
  holes on branes}},
  \href{https://doi.org/10.1088/1126-6708/2000/01/007}{\emph{JHEP} {\bfseries
  01} (2000) 007} [\href{https://arxiv.org/abs/hep-th/9911043}{{\ttfamily
  hep-th/9911043}}].

\bibitem{Emparan:1999fd}
R.~Emparan, G.T.~Horowitz and R.C.~Myers, \emph{{Exact description of black
  holes on branes. 2. Comparison with BTZ black holes and black strings}},
  \href{https://doi.org/10.1088/1126-6708/2000/01/021}{\emph{JHEP} {\bfseries
  01} (2000) 021} [\href{https://arxiv.org/abs/hep-th/9912135}{{\ttfamily
  hep-th/9912135}}].

\bibitem{Emparan:2002px}
R.~Emparan, A.~Fabbri and N.~Kaloper, \emph{{Quantum black holes as holograms
  in AdS brane worlds}},
  \href{https://doi.org/10.1088/1126-6708/2002/08/043}{\emph{JHEP} {\bfseries
  08} (2002) 043} [\href{https://arxiv.org/abs/hep-th/0206155}{{\ttfamily
  hep-th/0206155}}].

\bibitem{Emparan:2020znc}
R.~Emparan, A.M.~Frassino and B.~Way, \emph{{Quantum BTZ black hole}},
  \href{https://doi.org/10.1007/JHEP11(2020)137}{\emph{JHEP} {\bfseries 11}
  (2020) 137} [\href{https://arxiv.org/abs/2007.15999}{{\ttfamily
  2007.15999}}].

\bibitem{Plebanski:1976gy}
J.F.~Plebanski and M.~Demianski, \emph{{Rotating, charged, and uniformly
  accelerating mass in general relativity}},
  \href{https://doi.org/10.1016/0003-4916(76)90240-2}{\emph{Annals Phys.}
  {\bfseries 98} (1976) 98}.

\bibitem{Kolanowski:2023hvh}
M.~Kolanowski and M.~Toma\v{s}evi\'c, \emph{{Singularities in 2D and 3D quantum
  black holes}}, \href{https://doi.org/10.1007/JHEP12(2023)102}{\emph{JHEP}
  {\bfseries 12} (2023) 102}
  [\href{https://arxiv.org/abs/2310.06014}{{\ttfamily 2310.06014}}].

\bibitem{Panella:2024sor}
E.~Panella, J.F.~Pedraza and A.~Svesko, \emph{{Three-Dimensional Quantum Black
  Holes: A Primer}},
  \href{https://doi.org/10.3390/universe10090358}{\emph{Universe} {\bfseries
  10} (2024) 358} [\href{https://arxiv.org/abs/2407.03410}{{\ttfamily
  2407.03410}}].

\bibitem{Visser:1996iw}
M.~Visser, \emph{{Gravitational vacuum polarization. 1: Energy conditions in
  the Hartle-Hawking vacuum}},
  \href{https://doi.org/10.1103/PhysRevD.54.5103}{\emph{Phys. Rev. D}
  {\bfseries 54} (1996) 5103}
  [\href{https://arxiv.org/abs/gr-qc/9604007}{{\ttfamily gr-qc/9604007}}].

\bibitem{Visser:1996iv}
M.~Visser, \emph{{Gravitational vacuum polarization. 2: Energy conditions in
  the Boulware vacuum}},
  \href{https://doi.org/10.1103/PhysRevD.54.5116}{\emph{Phys. Rev. D}
  {\bfseries 54} (1996) 5116}
  [\href{https://arxiv.org/abs/gr-qc/9604008}{{\ttfamily gr-qc/9604008}}].

\bibitem{Visser:1997sd}
M.~Visser, \emph{{Gravitational vacuum polarization. 4: Energy conditions in
  the Unruh vacuum}},
  \href{https://doi.org/10.1103/PhysRevD.56.936}{\emph{Phys. Rev. D} {\bfseries
  56} (1997) 936} [\href{https://arxiv.org/abs/gr-qc/9703001}{{\ttfamily
  gr-qc/9703001}}].

\bibitem{Levi:2016quh}
A.~Levi and A.~Ori, \emph{{Versatile method for renormalized stress-energy
  computation in black-hole spacetimes}},
  \href{https://doi.org/10.1103/PhysRevLett.117.231101}{\emph{Phys. Rev. Lett.}
  {\bfseries 117} (2016) 231101}
  [\href{https://arxiv.org/abs/1608.03806}{{\ttfamily 1608.03806}}].

\bibitem{Christensen:1977jc}
S.M.~Christensen and S.A.~Fulling, \emph{{Trace Anomalies and the Hawking
  Effect}}, \href{https://doi.org/10.1103/PhysRevD.15.2088}{\emph{Phys. Rev. D}
  {\bfseries 15} (1977) 2088}.

\bibitem{Kontou:2023ntd}
E.-A.~Kontou and V.~Sacchi, \emph{{A generalization of the Hawking black hole
  area theorem}}, \href{https://doi.org/10.1007/s10714-024-03245-5}{\emph{Gen.
  Rel. Grav.} {\bfseries 56} (2024) 62}
  [\href{https://arxiv.org/abs/2303.06788}{{\ttfamily 2303.06788}}].

\bibitem{Hawking:1982dh}
S.W.~Hawking and D.N.~Page, \emph{{Thermodynamics of Black Holes in anti-De
  Sitter Space}}, \href{https://doi.org/10.1007/BF01208266}{\emph{Commun. Math.
  Phys.} {\bfseries 87} (1983) 577}.

\bibitem{Socolovsky:2017nff}
M.~Socolovsky, \emph{{Schwarzschild Black Hole in Anti-De Sitter Space}},
  \href{https://doi.org/10.1007/s00006-018-0822-6}{\emph{Adv. Appl. Clifford
  Algebras} {\bfseries 28} (2018) 18}
  [\href{https://arxiv.org/abs/1711.02744}{{\ttfamily 1711.02744}}].

\bibitem{Parikh2000}
M.K.~Parikh and F.~Wilczek, \emph{Hawking radiation as tunneling},
  \href{https://doi.org/10.1103/physrevlett.85.5042}{\emph{Physical Review
  Letters} {\bfseries 85} (2000) 5042}.

\end{thebibliography}\endgroup

\end{document}